\documentclass{article}
\usepackage{amsmath}
\usepackage{amssymb}
\def\dv{\mathbb}
\def\0{\mbox{\sf 0}}
\def\1{\mbox{\sf 1}}
\def\2{\mbox{\sf 2}}
\def\3{\mbox{\sf 3}}
\def\boP{\mbox{\boldmath $P$}}
\def\sfX{\mbox{\sf X}}
\def\isfX{\mbox{\scriptsize\sf X}}
\def\sfY{\mbox{\sf Y}}
\def\sfZ{\mbox{\sf Z}}
\def\bxi{\mbox{\boldmath$\xi$}}
\def\ibxi{\mbox{\scriptsize\boldmath$\xi$}}
\def\ci{\mbox{$\perp\!\!\!\perp\,$}}
\def\ici{\mbox{\footnotesize $\perp\!\!\!\perp\,$}}
\def\notci{\mbox{$\not\perp\!\!\!\perp$}}
\def\ingl{\square\,}
\def\diff{\Delta\,}
\newtheorem{proposition}{Proposition}
\newtheorem{lemma}{Lemma}
\newtheorem{cor}[lemma]{Corollary}
\newtheorem{remark}{Remark}
\newtheorem{theorem}{Theorem}
\newtheorem{example}{Example}
\newtheorem{open}{Open question}
\newenvironment{proof}{\begin{trivlist}\item[] \mbox{\it Proof. }}
{\hfill$\Box$ \end{trivlist}}

\title{Conditional independence structures over four discrete random variables revisited:\\
conditional Ingleton inequalities \thanks{This work has been submitted to the IEEE for possible publication. Copyright may be transferred without notice, after which this
version may no longer be accessible.}}
\author{by Milan Studen\'{y}}

\date{\today \\[1ex] {\em dedicated to the memory of Franti\v{s}ek Mat\'{u}\v{s}} }

\begin{document}

\maketitle

\begin{abstract}
The paper deals with linear
information inequalities valid for entropy functions induced by discrete random variables. Specifically, the so-called {\em conditional Ingleton inequalities\/}  are in the center of interest: these are valid under conditional independence assumptions
on the inducing \mbox{random} variables.
We discuss five inequalities of this particular type, four of which has appeared earlier in the literature.
Besides the proof of the new fifth inequality, simpler proofs of (some of) former inequalities are presented.
These five information inequalities are used to characterize all conditional independence structures
induced by four discrete random variables.
\end{abstract}

\section{Introduction}\label{sec.introduction}
The motivation for this paper is explained and its structure is described.

\subsection{Motivation}\label{ssec.motivation}
The concept of {\em conditional independence\/} (CI) has been studied in probability theory \cite[\S\,25.3]{Loe55}
and statistics \cite{Daw79JRSS,MouRol84S} for many years. This concept has a key theoretical role in 
probabilistic reasoning \cite{Pea88} and graphical statistical models \cite{Lau96}. This is because the implications
between CI statements are interpreted in this context as substantial abstract properties of (= ``axioms" for)
the respective (probabilistic) conditional irrelevance relation. The first attempts to describe abstract properties of
probabilistic CI in terms of (a finite number of) elementary CI implications occurred in the end of the 1970s \cite{Daw79JRSS,Spo80JPL}.
It was shown in the 1990s that probabilistic CI structures cannot be characterized by means of a finite number of
CI implications of this kind \cite{Stu92conf} while some important substructures of probabilistic CI structures
can be characterized in this way \cite{Mat94TCS}.

One of the goals motivated by this idea was to characterize CI \mbox{structures} induced by four discrete random
variables. Note in this context that, for a theoretical reason, the CI structures over a fixed number of 
random variables can be characterized by means of a finite number of such elementary CI implications but the number
of these CI implications (= ``axioms" for CI) grows rapidly with the number of random variables.
The goal has been achieved in the end of 1990s by Mat\'{u}\v{s} in a series of three papers \cite{MatStu95CPC,Mat95CPC,Mat99CPC}; the author of this article was a co-author of the first paper in the series.
The final solution was the result of Mat\'{u}\v{s}'s enormous efforts: it took him more than five years to reach his goal. Note in this context that the number of CI structures over four discrete random variables is very high, in fact, one has 18\,478 standard such CI structures.

To prove the validity of elementary CI implications (= of ``axioms" for CI) Mat\'{u}\v{s} \cite{Mat95CPC,Mat99CPC} used a number of methods that he has developed. Some of the CI implications were derived as consequences
of considerations within a wider framework of $\sigma$-algebras, some of them followed from non-negativity of the
Kullback-Leibler divergence of specially constructed probability distributions and a few of them were based on the characterization of CI in terms of factorization of probability density. In the last paper in the series
two remaining CI implications were derived as consequences of conditional information inequalities.
That last method was inspired by then recent paper by Zhang and Yeung \cite{ZhaYeu97ITI} in which they
revealed the first (non-Shannon) conditional information inequality;  Mat\'{u}\v{s} in \cite{Mat99CPC} proved the second
conditional inequality. Thus, one can say, with a little exaggeration, that each of the found CI implications from \cite{Mat95CPC,Mat99CPC} has its own specific proof. The interest in conditional information inequalities has been restored some years later when Kaced and Romashchenko \cite{KacRom13IEEE} found other such inequalities and introduced the ways to classify them.

The source of motivation for this paper was the following simple question: could it be the case that
all the CI implications that were revealed in case of four discrete random variables
can be derived using only one procedure? The surprising answer is ``yes": it appears that
all these CI implications can really be derived as consequences of conditional information
inequalities valid for (entropy functions induced by) discrete random variables.
These inequalities are special in a certain sense: they can be viewed as {\em conditional Ingleton inequalities\/}
and only five of them are enough to cover all 19 necessary CI implications.
Note in this context that there are also further valid CI implications/equivalences which, however,
can be verified even easier.
Another remark is that this paper, unlike the original series of papers \cite{MatStu95CPC,Mat95CPC,Mat99CPC},
deals with standard CI structures, which means that the functional dependencies among
the considered random variables are not taken into consideration.
This is not a substantial difference because every augmented CI structure (= the one with
functional dependence statements involved) can be viewed as an extension of a unique
standard CI structure (= the one without functional dependencies).

The contribution of this paper is a new (= the fifth) conditional Ingleton inequality,
simpler proofs of two former such inequalities from \cite{Mat99CPC,KacRom13IEEE}
and the derivation of all elementary CI implications over four discrete random variables
using these five conditional Ingleton inequalities. For sake of completeness we also recall,
or perhaps reinterpret, the proofs of two remaining inequalities so that the reader interested
in the characterization of CI structures over four discrete random variables has a complete proof at disposal.
Thus, altogether, we present a simpler proof of the characterization of CI structures over four
variables than was the original one from \cite{MatStu95CPC,Mat95CPC,Mat99CPC}.
There was also another conjectured conditional Ingleton inequality (= the sixth one), whose validity would not have been
in contradiction with the validity of 19 above-mentioned elementary CI implications, but that one appeared
not to be true. Thus, a counter-example to its validity is also given.

\subsection{Structure of the paper}\label{ssec.content}
Basic concepts of the paper are introduced in Section~\ref{sec.preli}. We formally define the concepts of CI,
entropy function, polymatroid, and Ingleton inequality there; we also recall a few fundamental facts used later.
Section~\ref{sec.history} then provides the reader with an overview of related former results; it can be viewed
as a kind of extended introduction and can be skipped without losing understanding of the rest of the paper.
Specifically, we recall the results on characterization of CI structures
over 4 discrete \mbox{random} variables (in more details), on representations of polymatroids, and on information inequalities.
Section~\ref{sec.condi-Ingleton} is then
devoted particularly to conditional Ingleton inequalities. We present 5 inequalities of this type there,
whose proofs, however, are moved to Appendix (Section~\ref{sec.appendix-proofs}).
Additionally, five counter-examples to the (probabilistic) validity of the Ingleton \mbox{inequality} are given
in Section~\ref{sec.condi-Ingleton} 
and a few open tasks are formulated there. Section~\ref{sec.CI-structures} then brings an elegant application of the results
from Section~\ref{sec.condi-Ingleton}: we characterize all (standard) CI structures induced by
four discrete random variables there.
In Conclusions (Section~\ref{sec.conclusions}) we comment further potential applications
of the presented results.

\section{Preliminaries}\label{sec.preli}
Throughout the paper a finite non-empty {\em basic set\/ $N$\/}
will be considered; its elements will index (= correspond to) discrete random variables.
To avoid the trivial case we assume $|N|\geq 2$.
Shortened notation for some subsets of $N$ will often be used:
union of two subsets $X,Y\subseteq N$ may be denoted by a juxtaposition of respective symbols: $XY:= X\cup Y$. Analogously, the symbol for an element $i\in N$ will also serve to denote the respective singleton subset of $N$: $i:=\{i\}$.
The power set of $N$ will be denoted by ${\cal P}(N):=\{S:S\subseteq N\}$.

\subsection{Discrete random vector over a basic set}\label{ssec.random-var}
A {\em discrete random variable\/} is a measurable function $\xi$
from a probability space $(\Omega,{\cal A},\boP)$ to a non-empty finite set $\sfX$, called
the {\em sample space\/} for $\xi$. The {\em density\/} (for $\xi$) is then a function $p$ on $\sfX$ defined by
$p(x):= \boP (\{\omega\in\Omega\,:\ \xi(\omega)=x\,\})$ for $x\in\sfX$.
Note that the probability space is only an auxiliary tool here, the object of interest
is, in fact, the induced probability distribution on the sample space and this object is fully
described by the density. Observe that $p$ is a density (for a random variable with
sample space $\sfX$) iff $p:\sfX\to [0,1]$ and $\sum_{x\in\isfX} p(x)=1$.

By a {\em discrete random vector\/} over $N$ will be understood an indexed collection $\bxi=[\xi_{i}]_{i\in N}$
of random variables on a shared probability space $(\Omega,{\cal A},\boP )$ with each $\xi_{i}$ taking values in
its {\em individual\/} non-empty finite {\em sample space $\sfX_{i}$}.

Given such a random vector $\bxi$ over $N$, we introduce a symbol
$\sfX_{A}:=\prod_{i\in A} \sfX_{i}$ for any $\emptyset\neq A\subseteq N$,
to denote the set of all possible {\em configurations\/} of values for the (random) sub-vector $\xi_{A}:=[\xi_{i}]_{i\in A}$.
Each sub-vector $\xi_{A}$ of $\bxi$ can thus be interpreted as a random variable with
sample space $\sfX_{A}$; the random vector $\bxi$ itself can thus be viewed as a single random
variable as well, with values in the {\em joint sample space $\sfX_{N}$}.
Given a configuration $x\in\sfX_{A}$ and $i\in A$, the symbol $x_{i}$ will denote the respective component of $x$, that is,
$x=[x_{i}]_{i\in A}$, where $x_{i}\in\sfX_{i}$.

The {\em marginal density\/} for $A\subseteq N$ is defined by the formula
$$
p_{A}(x) := \sum\,  \{\, p(y)\,: \ y\in\sfX_{N} ~\mbox{\rm satisfying}~ x_{i}=y_{i}~~ \mbox{for any $i\in A$}\,\}\quad \mbox{for $x\in\sfX_{N}$.}
$$
We have intentionally introduced it as a function on $\sfX_{N}$ depending on the components from $A$.
The special cases are then the {\em joint density $p_{N}$} and the density $p_{\emptyset}$ for the empty set, which
is a constant function on $\sfX_{N}$ taking the value $1$.
Nevertheless, if appropriate, one can view the marginal density $p_{A}$ as a function on
the respective marginal sample $\sfX_{A}$ (provided $A\neq \emptyset$).

\subsection{Probabilistic conditional independence}\label{ssec.prob-CI}
Probabilistic CI structure induced by a random vector $\bxi$ over $N$ is a discrete mathematical structure describing
stochastic independence (and dependence) relations among (random) sub-vectors of $\bxi$.
Having three subsets $X,Y,Z\subseteq N$ of the basic set we say
that $\xi_{X}$ is {\em conditionally independent\/} of $\xi_{Y}$ {\em given\/} $\xi_{Z}$ and write $\xi_{X}\ci\xi_{Y}\,|\,\xi_{Z}$ if
\begin{equation}
\forall\, x\in\sfX_{N}\qquad
p_{XYZ}(x)\cdot p_{Z}(x) = p_{XZ}(x)\cdot p_{YZ}(x)\,.
\label{eq.CI-def}
\end{equation}
An alternative notation is $X\ci Y\,|\,Z\,\,[\bxi]$. The respective ordered triplet of sets
will also be denoted by $(X,Y|Z)$; we use a bar to separate the third component interpreted
as the conditioning variable set $Z$.

Note that the definition \eqref{eq.CI-def} works for any triplet of sets $X,Y,Z$ although these three
sets are typically assumed to be pairwise disjoint.
The case \mbox{$Z=\emptyset$} corresponds to classic (unconditional) stochastic independence, denoted also by $\xi_{X}\ci\xi_{Y}$,
alternatively by $X\ci Y\,\,[\bxi]$. The case $Y=X$ corresponds to functional dependence of $\xi_{X}$ on $\xi_{Z}$:
the reader can easily deduce from \eqref{eq.CI-def} that $\xi_{X}\ci\xi_{X}|\xi_{Z}$
is equivalent to the existence of a function $f:\sfX_{Z}\to\sfX_{X}$ such that $\xi_{X}=f(\xi_{Z})$.

There is a number of equivalent definitions of probabilistic CI, one of them in terms of {\em conditional density},
defined by the ratio $p_{X|Z}(x):=\frac{p_{XZ}(x)}{p_{Z}(x)}$ for $x\in\sfX_{N}$ provided that $p_{Z}(x)>0$,
with disjoint $X,Z\subseteq N$. Thus, the reader can verify easily that,
in case of pairwise disjoint $X,Y,Z\subseteq N$ with $X\neq\emptyset\neq Y$,
one has $\xi_{X}\ci\xi_{Y}\,|\,\xi_{Z}$ iff $p_{X|YZ}(x)=p_{X|Z}(x)$ for each $x\in\sfX_{N}$ such that $p_{YZ}(x)>0$.
To say that in yet another way, one has $\xi_{X}\ci\xi_{Y}\,|\,\xi_{Z}$ iff the conditional density $p_{X|YZ}(x)$
does not depend on the (value of a) sub-configuration for $Y$. %
In binary case, that is, in case $\sfX_{i}=\{\0,\1\}$ for $i\in N$, the following
{\em cross-product criterion\/} can be used: if $|X|=1=|Y|$ and $q$ is the marginal density
for $XYZ$ interpreted as a function on $\sfX_{XYZ}$ then $\xi_{X}\ci\xi_{Y}\,|\,\xi_{Z}$ iff
$q(\0,\0,z)\cdot q(\1,\1,z)= q(\0,\1,z)\cdot q(\1,\0,z)$ for any
configuration $z\in\sfX_{Z}$ for $Z$.

\smallskip
Formally, the CI structure induced by a random vector $\bxi$ over $N$ is a
certain ternary relation on the power set ${\cal P}(N)$.
Specifically, the {\em augmented CI structure\/} induced by a random vector $\bxi$ over $N$ is
$$
\{\, (X,Y|Z)\in {\cal P}(N)\times{\cal P}(N)\times {\cal P}(N) \,:\ \xi_{X}\ci\xi_{Y}|\xi_{Z}\,\}\,.
$$
The {\em standard CI structure\/} is the above ternary relation on ${\cal P}(N)$
confined to triplets of pairwise disjoint subsets of $N$ as required traditionally in \cite{Pea88,Lau96}.
Note that discrete probabilistic CI structures over $N$ are naturally ordered by inclusion
of involved independence triplets (for further details see Section~\ref{ssec.lattice}).

\subsection{Conditional product of distributions}\label{ssec.cond-prod}
A discrete {\em probability distribution over $N$} is simply the distribution of a discrete
random vector over $N$. Put it in another way: a discrete distribution \mbox{$P$ over $N$} is specified by
a collection of individual non-empty finite sample spaces $\sfX_{i}$, $i\in N$, and by its density
$p:\sfX_{N}\to [0,1]$ defined on the respective joint sample space.
Given $\emptyset\neq A\subset N$, the {\em marginal\/} (distribution) of $P$ for $A$ can be introduced as the
distribution of the random sub-vector $\xi_{A}$. Thus, it is a probability distribution over $A$
with individual sample spaces $\sfX_{i}$, $i\in A$, inherited from $P$. The marginal density
can be computed from the joint one by summing over configurations in $\sfX_{N\setminus A}$.

Given pairwise disjoint sets $A,B,C\subseteq N$ with $A\neq\emptyset\neq B$, we say that (discrete) probability distributions $Q$ over $AC$ and $R$ over $BC$ are {\em consonant\/} if the marginals of $Q$ and $R$ for $C$ coincide,
which involves the requirement that the individual sample spaces for $i\in C$ coincide. Given
such a pair of consonant distributions their {\em conditional product\/} is a (discrete) probability
distribution $P$ over $ABC$ with individual sample spaces inherited from $Q$ and $R$ whose density $p$ on $\sfX_{ABC}$
is defined by the following formula:
$$
p(a,b,c) ~:=~
\left\{
\begin{array}{cl}
\,\frac{q(a,c)\cdot r(b,c)}{r(c)}
& ~~\mbox{if $p(c)>0$,}\\[2ex]
\,0 & ~~\mbox{otherwise,}
\end{array}
\right.
$$
where $a,b,c$ belong to the respective marginal sample spaces and $q(a,c)$ denotes
the value of the marginal density of $Q$ for $AC$ in the configuration $[a,c]$. Easy
observations are that the marginals of the conditional product $P$ are the given distributions $Q$
and $R$ and one, moreover, has $\xi_{A}\ci \xi_{B}\,|\,\xi_{C}$ for a random vector $\bxi$ over $ABC$
having the conditional product $P$ as its distribution.

\subsection{Semi-graphoids}\label{ssec.semi-graph}
In this paper we also deal with (standard) {\em abstract CI structures\/} over $N$ which are simply
ternary relations on ${\cal P}(N)$ (confined to triplets of pairwise disjoint sets). A convention
is accepted that ordered triplets $(X,Y|Z)$ of sets involved in such ternary relations are interpreted as
abstract independence statements, which is expressed by notation $X\ci Y\,|\,Z$.

Abstract CI structures can be induced by various mathematical objects, not just by random
vectors. Nevertheless, they typically comply with some basic formal properties of (= ``axioms" for)
conditional irrelevance relations which were already pinpointed in \cite{Daw79JRSS}.
Thus, following to \cite{Pea88}, we will say that a standard abstract CI structure over $N$ is a {\em semi-graphoid\/}
if it satisfies the next three conditions: given pairwise disjoint sets $X,Y,Z,U\subseteq N$, one has
\begin{itemize}
\item[(S:0)] \,$\emptyset\,\ci Y\,|\,Z$,
\item[(S:1)] $X\ci Y\,|\,Z ~ \Leftrightarrow~ Y\ci X\,|\,Z$,
\item[(S:2)] $X\ci YZ\,|\,U ~~ \Leftrightarrow~~ [\,X\ci Y\,|\,ZU ~\&~ X\ci Z\,|\,U\,]$\,.
\end{itemize}
One can analogously introduce semi-graphoids in context of augmented abstract CI structures:
admit intersecting sets in (S:1) and (S:2) and replace (S:0) by
\begin{itemize}
\item $X\subseteq Z ~ \Rightarrow~ X\ci Y\,|\,Z$\,.
\end{itemize}
The fact that every probabilistic CI structure is a semi-graphoid easily follows
from later observations (see Lemma~\ref{lem.entrop-polym} in Section~\ref{ssec.entropy}).
\smallskip

The semi-graphoid properties imply that one need not retain
(information about) all CI statements in order to represent a CI structure.
A triplet $(X,Y|Z)$ of pairwise disjoint subsets of $N$
will be called {\em elementary\/} if $|X|=1=|Y|$. The point is that elementary triplets
are enough to represent a semi-graphoid.

\begin{lemma}\rm \label{lem.local-semi}
Given a semi-graphoid over $N$ (denoted using the symbol $\ci$) and a triplet $(X,Y|Z)$ of pairwise disjoint subsets of $N$ one has
\begin{equation}
X\ci Y\,|\, Z  ~\Leftrightarrow~ [\,\forall\, i\in X,~j\in Y,~K\,:\,Z\subseteq K\subseteq XYZ\setminus\{i,j\},
\quad  i\ci j\,|\,K \,\,]\,.
\label{eq.local}
\end{equation}
\end{lemma}

\begin{proof}
We leave it to the reader to verify \eqref{eq.local} by induction on the cardinality of $XY$.
The induction premise follows immediately from (S:0) if $|XY|\leq 2$ while the induction step can be verified
using (S:2).
\end{proof}

Taking into consideration the symmetry property (S:1) one can deduce
that to represent a standard semi-graphoid over $N$, $|N|=n$, in the memory of a computer it is enough to have
$n\cdot (n-1)\cdot 2^{n-3}$ bits, each of which corresponds to a pair
of 
mutually equivalent elementary triplets $(i,j|K)$ and $(j,i|K)$.

In the context of augmented abstract CI structures one has, however, to introduce additional
(non-disjoint) elementary triplets $(i,i|N\setminus i)$ for $i\in N$; Lemma~\ref{lem.local-semi}
can then be extended to that case as well.


\subsection{Polymatroids}\label{ssec.polymatroid}
A central role in our exposition will have real set functions $h:{\cal P}(N)\to {\dv R}$.
Given such a function $h$ and sets $X,Y,Z\subseteq N$ we introduce a special shorthand
\begin{equation}
\diff h(X,Y|Z) ~:=~ h(XZ)+h(YZ)-h(XYZ)-h(Z)
\label{eq.differ}
\end{equation}
to denote the corresponding {\em difference expression\/} for the ordered triplet $(X,Y|Z)$ of subsets of $N$.
The symbol of the function $h$ will sometimes be omitted; we will do so in situations
when we wish to emphasize that \eqref{eq.differ} defines a functional $\diff (X,Y|Z):~ h\mapsto \diff h(X,Y|Z)$
on ${\dv R}^{{\cal P}(N)}$.
Typically, the considered triplet $(X,Y|Z)$ of subsets of $N$ will be composed of pairwise
disjoint sets and, in this case, it will be associated with an abstract CI statement $X\ci Y\,|\,Z$.
\smallskip

The rank function of a {\em polymatroid\/} is a real set functions $h:{\cal P}(N)\to {\dv R}$
which satisfies $h(\emptyset)=0$ and $0\leq \diff h(X,Y|Z)$ for any triplet $X,Y,Z\subseteq N$ of
subsets of the basic set. In case $I\subseteq J\subseteq N$ and $(X,Y|Z)=(J,J|I)$ one has
$0\leq \diff h(J,J|I)= h(J)-h(I)$, which implies that every rank function $h$ of a polymatroid is
{\em non-decreasing}, and, therefore, non-negative.
The requirement $0\leq \diff h(X,Y|Z)$ for any triplet $(X,Y|Z)$ of pairwise disjoint subsets of $N$ then means
that $h$ is {\em submodular}, that is, $h(I)+h(J)\geq h(I\cup J)+ h(I\cap J)$ for any $I,J\subseteq N$.
Conversely, given three arbitrary sets $X,Y,Z\subseteq N$, the equality
$$
\diff (X,Y|Z) ~=~ \diff(A,BC | Z\,) + \diff(B,C | AZ\,) + \diff(C,C | ABZ\,)\,,
$$
with $A:=X\setminus YZ$, $B:=Y\setminus XZ$, $C:= (X\cap Y)\setminus Z$, allows
one to show that any non-decreasing submodular function $h$ satisfying $h(\emptyset)=0$ defines
a polymatroid; this relates our definition above to the common one from \cite[\S\,2.2]{Fuj91}.
The concept itself can be viewed as a generalization
of the concept of a matroid.
Indeed, by \cite[Corollary~1.3.4]{Oxl11}, an integer-valued set functions
$h:{\cal P}(N)\to {\dv Z}$ is known to be a rank function of a {\em matroid\/} if it is non-decreasing,
submodular and bounded by cardinality: if $I\subseteq N$ then $0\leq h(I)\leq |I|$.
\smallskip

Polymatroids can also be assigned abstract CI structures. Specifically, given a rank function
$h:{\cal P}(N)\to {\dv R}$ of a polymatroid (over $N$) we define
$$
X\ci Y\,|\,Z\,\,[h] ~:=~ \diff h(X,Y|Z)=0\quad \mbox{for any triplet $(X,Y|Z)$ of subsets of $N$.}
$$
The following observation easily follows from the definition.

\begin{lemma}\rm \label{lem.polym-semig}
The CI structure induced by a polymatroid is a semi-graphoid.
\end{lemma}

\begin{proof}
The fact $\diff (\emptyset,Y|Z)=0$ gives (S:0) while
$\diff (X,Y|Z)=\diff (Y,X|Z)$ gives (S:1). Given a rank function $h:{\cal P}(N)\to {\dv R}$
of a polymatroid, the relation $\diff h(X,YZ|U)=\diff h(X,Y|ZU) +\diff h(X,Z|U)$ together
with non-negativity of the involved terms implies the property (S:2).
\end{proof}

\subsection{Entropy function}\label{ssec.entropy}
Given a discrete random variable $\xi$ with a sample space $\sfX$, whose distribution
is given by a density $p:\sfX\to [0,1]$, its {\em entropy\/} $H(\xi)$ is given by the formula
$$
H(\xi) ~:=~ -\sum_{x\in\isfX: p(x)>0}\, p(x)\cdot \ln (p(x))\,.
$$
It is clearly a non-negative real number. Thus, every discrete random vector $\bxi=[\xi_{i}]_{i\in N}$ can be assigned its
{\em entropy function\/} $h_{\ibxi}:{\cal P}(N)\to [0,\infty)$ defined by
$$
h_{\ibxi}(A) ~:=~ H(\xi_{A})\qquad \mbox{for every $A\subseteq N$.}
$$

\begin{lemma}\rm \label{lem.entrop-polym}
Given a discrete random vector $\bxi$ over $N$, the entropy function $h_{\ibxi}$ is the rank function of a
polymatroid (over $N$). Moreover, for every ordered triplet $(X,Y|Z)$ of subsets of $N$ one has
\begin{equation}
\xi_{X}\ci\xi_{Y}|\xi_{Z}\quad\Leftrightarrow\quad \diff h_{\ibxi}(X,Y|Z)=0 \quad\Leftrightarrow\quad
X\ci Y\,|\,Z\,\,[h_{\ibxi}]\,.
\label{eq.CI-entrop}
\end{equation}
In particular, the CI structure induced by $\bxi$ is a semi-graphoid.
\end{lemma}

\begin{proof}
Elementary information-theoretical inequalities, see
\cite[Appendix~14.A]{Yeu08}, \mbox{imply} that the entropy function $h_{\ibxi}$ is a rank function of a polymatroid.
The relation \eqref{eq.CI-entrop} then follows from the fact that the value $\diff h_{\ibxi}(X,Y|Z)$
is exactly the so-called {\em conditional mutual information\/} between $\xi_{X}$ and $\xi_{Y}$ given $\xi_{Z}$.
This quantity is known to vanish just in case of the validity of the CI statement;
see \cite[Theorem~2.34]{Yeu08} or \cite[Corollary~2.2]{Stu05}.
The rest follows from Lemma~\ref{lem.polym-semig}.
\end{proof}

Given discrete probability distributions $Q$ and $R$ on some sample space $\sfX$,
with respective densities $q$ and $r$, we say
that $Q$ is {\em absolutely continuous\/} with respect to $R$ if\/ $r(x)=0$ implies $q(x)=0$
for any $x\in\sfX$. We might alternatively say that {\em $R$ dominates $Q$} or that $Q$ is dominated by $R$.
If this is the case then one can define the {\em Kullback-Leibler divergence\/} of $Q$ with respect
to $R$ by the formula
$$
D(Q\,\|\,R) ~:=~ \sum_{x\in\isfX: q(x)>0}\, q(x)\cdot \ln \frac{q(x)}{r(x)}\,.
$$
Well-known facts are that $D(Q\,\|\,R)\geq 0$ and the equality holds iff $Q=R$ \cite[Theorem~2.31]{Yeu08}.
Observe that, given a probability distribution $P$ over $N$ and pairwise disjoint sets $A,B,C\subseteq N$ with $A\neq\emptyset\neq B$, the conditional product $\overline{P}$ of the marginals of $P$ for $AC$ and $BC$
dominates the (original) distribution $P$. The Kullback-Leibler divergence $D(P\,\|\,\overline{P})$ is then
the respective conditional mutual information.

\subsection{Ingleton expression in different forms}\label{ssec.Ingleton}
Another important quantity exploited in this paper is the expression defining the so-called
Ingleton inequality, discussed in more details in Section \ref{ssec.CI4}. 
We are going to call that quantity (the) {\em Ingleton expression\/} and introduce a \mbox{special} symbol
for it, following the notation from \cite{MatStu95CPC}.

Given a real set functions $h:{\cal P}(N)\to {\dv R}$ and four {\em pairwise disjoint\/}
subsets $X,Y,Z,U\subseteq N$ of the basic set $N$, we introduce another special shorthand
\begin{eqnarray}
\lefteqn{\hspace*{-32mm}\ingl h(X,Y) \,:=\, -\,h(XY)+h(XZ)+h(XU)+h(YZ)+h(YU)+h(ZU)}\label{eq.Inglet}\\
&& \hspace*{-10mm} -\,h(Z)-h(U)-h(XZU)-h(YZU) \nonumber
\end{eqnarray}
to denote the respective Ingleton expression for $(X,Y|Z,U)$. Note that a more appropriate
notation would be $\ingl h(X,Y|Z,U)$ because the expression \eqref{eq.Inglet} does depend on sets $Z$ and $U$. Nevertheless,
we will solely use this notation in a situation when the sets $X,Y,Z,U\subseteq N$ are fixed and no misunderstanding
can occur. Observe that \eqref{eq.Inglet} is invariant relative to mutual exchanges $X\leftrightarrow Y$ and
$Z\leftrightarrow U$. Like in case of the difference expression from
Section~\ref{ssec.polymatroid} the symbol of the function $h$ might be omitted in order to emphasize that \eqref{eq.Inglet}
defines a functional $\ingl (X,Y):~ h\mapsto \ingl h(X,Y|Z,U)$
on ${\dv R}^{{\cal P}(N)}$.

The point is that the expression \eqref{eq.Inglet} has a number of different forms, which were named,
somewhat figuratively, ``masks" in \cite[\S\,4]{MatStu95CPC}. Note that only five ``masks" were
mentioned in \cite[\S\,4]{MatStu95CPC} and are recalled in Lemma~\ref{lem.Inglet-forms} below but thanks
to the invariance of \eqref{eq.Inglet} relative to
$X\leftrightarrow Y$ and $Z\leftrightarrow U$ one can recognize even 14 different forms of this
expression. In fact, besides those 14 basic forms, there are further possible re-writings which
have six difference terms.

\begin{lemma}\em \label{lem.Inglet-forms}
Given four pairwise disjoint subsets $X,Y,Z,U\subseteq N$ of the basic set the
Ingleton expression \eqref{eq.Inglet} has the following different forms:
\begin{description}
\item[\mbox{\rm (M.1)~\,}] \,$\ingl (X,Y) \,=\, \diff (Z,U|X) +  \diff (Z,U|Y)
+ \diff (X,Y|\emptyset) - \diff (Z,U|\emptyset)$
\item[\mbox{\rm (M.2)~\,}] \,$\ingl (X,Y) \,=\, \diff (Z,U|Y) +  \diff (X,Z|U)
+ \diff (X,Y|\emptyset) - \diff (X,Z|\emptyset)$
\item[\mbox{\rm (M.3)~\,}] \,$\ingl (X,Y) \,=\, \diff (X,Y|Z) +  \diff (X,Z|U)
+ \diff (Z,U|Y) - \diff (X,Z|Y)$
\item[\mbox{\rm (M.4)}] \,$\ingl (X,Y) \,=\, \diff (X,Y|Z) +  \diff (X,Y|U)
+ \diff (Z,U|XY) - \diff (X,Y|ZU)$
\item[\mbox{\rm (M.5)}] \,$\ingl (X,Y) \,=\, \diff (X,Y|Z) +  \diff (X,Z|U)
+ \diff (Z,U|XY) - \diff (X,Z|YU)$
\end{description}
\end{lemma}

\begin{proof}
Easy verification of every form is left to the reader:  substitute \eqref{eq.differ} into the
right-hand side of it and, after canceling some terms, obtain \eqref{eq.Inglet}.
\end{proof}

\subsection{Lattice of CI structures}\label{ssec.lattice}
A partially ordered set $({\cal Z},\preceq)$ is called a {\em lattice\/} \cite[\S\,I.4]{Bir95} if every two-element subset of ${\cal Z}$ has both the least upper bound, also named the supremum or the {\em join}, and the greatest lower bound, also named the infimum or the {\em meet}. A finite lattice is necessarily {\em complete\/} which means that the requirement above holds for any subset of ${\cal Z}$. One of standard examples of a finite lattice is the {\em face-lattice\/} of a polyhedral cone \cite[Theorem~2.7(v)]{Zie95}.
A lattice $({\cal Z},\preceq)$ is {\em anti-isomorphic\/} to a lattice
$({\cal Z}^{\prime},\preceq^{\prime})$ if there is a one-to-one mapping $\iota$ from ${\cal Z}$ onto ${\cal Z}^{\prime}$ which reverses the ordering: for $x,y\in {\cal Z}$, one has $x\preceq y$ iff $\iota(y)\preceq^{\prime}\iota(x)$.

An element $e$ in a lattice ${\cal Z}$ is called {\em meet-irreducible\/} if it
cannot be written as the infimum of two elements of ${\cal Z}$ which are both
different from $e$. In a finite lattice, every element can be written as
the infimum of a set of meet-irreducible elements.
Examples of meet-irreducible elements are the unique maximal element ${\bf 1}$ in the lattice and its
{\em sub-maximal\/} elements, called {\em co-atoms}.
These are the elements $c\in {\cal Z}$ distinct from ${\bf 1}$
such that the only elements $e\in {\cal Z}$ satisfying $c\preceq e$ are $e=c$ and $e={\bf 1}$.
On the other hand, a finite lattice can have additional meet-irreducible elements besides those
mentioned above.

Recall from Section~\ref{ssec.prob-CI} that discrete probabilistic CI structures over $N$
are naturally ordered {\em by inclusion\/} (of involved ordered triplets of subsets of $N$).
The point is that they form a complete lattice relative to this ordering. This is because
the {\em intersection\/} of two discrete probabilistic CI structures over $N$ is
a discrete probabilistic CI structure over $N$, which fact follows from the following construction.
Given densities $q:\sfY_{N}\to [0,1]$ and $r:\sfZ_{N}\to [0,1]$ of discrete probability distributions
$Q$ and $R$ over $N$ one can put
$$
p([y,z]) ~:=~ q(y)\cdot r(z)\qquad \mbox{for any $[y,z]\in \sfY_{N}\times\sfZ_{N}$,}
$$
which defines the density of a probability distribution $P$ over $N$ with individual sample spaces $\sfX_{i}:=\sfY_{i}\times\sfZ_{i}$ for any $i\in N$. The CI structure induced by $P$ appears
to be the intersection of CI structures induced by $Q$ and $R$.
The fact that CI structures are closed under intersection allows one to characterize
them by means of meet-irreducible elements in the lattice of these structures.

\section{Overview of related former results}\label{sec.history}
This section, recalling earlier findings relevant to the topic of study,
can be skipped without losing understanding of the rest of the paper.
To interpret geometrically some of the notions concepts from
polyhedral geometry are used that are not defined here; definitions
can be found in books on this topic \cite{Sch03,Zie95}.

\subsection{Representations of polymatroids}\label{ssec.repre-polym}
{\em Matroid theory\/} was introduced already in the 1930s \cite{Whi35AJM} as an abstract theory of independence  inspired by the concepts of independence emerging in linear algebra and other fields. There is a number of
crypto-equivalent definitions of the concept of a matroid in sense that the equivalence is not apparent (at first glance)
and some transitional constructions are needed; see \cite{Oxl11} or \cite[Chapter\,39]{Sch03} for details.
One of the equivalent definitions is in terms of the rank function (of a matroid) given in Section~\ref{ssec.polymatroid}. Some of the matroids are {\em linearly representable\/} over a field ${\dv F}$,
which means that they are defined through linear independence relationships within a finite collection of vectors from a linear space over ${\dv F}$.

The concept of a {\em polymatroid\/} was introduced by Edmonds \cite{Edm70} in the early 1970s as a certain bounded polyhedron in the space ${\dv R}^{N}$ whose defining inequalities correspond to subsets of $N$.
Tight upper bounds for the inequalities then define the values of the respective rank function,
determining the polyhedron uniquely. The terminology was motivated by the fact that this concept
generalizes the concept of an ``independent set polytope" for a matroid \cite[\S\,40.2]{Sch03}.
Later authors dealing with polymatroids \cite[\S\,2.2]{Fuj91}, however, found it to be more appropriate to define polymatroids in terms of their rank functions instead; we have done so
in Section~\ref{ssec.polymatroid} as well. One can extend the concept of linear representability
(from matroids) to polymatroids whose rank functions are integer-valued \cite[p.\,30]{Fuj91}.

One can distinguish different forms of probabilistic representability for a polymatroid;
most of definitions below were inspired by Mat\'{u}\v{s} \cite{Mat94IJGS,Mat07IEEE}.
Let us say that a polymatroid over $N$ with a rank function $h\in {\dv R}^{{\cal P}(N)}$ is
\begin{itemize}
\item {\em strongly probabilistically representable\/} if there exists a constant $c>0$ and
a discrete random vector $\bxi$ over $N$ such that $c\cdot h(S)=h_{\ibxi}(S)$ for any $S\subseteq N$,
\item (strongly) {\em asymptotically probabilistically representable\/} if there exists a constant $c>0$ and
a sequence of discrete random vectors $\bxi_{n}$ over $N$ such that $c\cdot h(S)= \lim_{n\to\infty} h_{\ibxi_{n}}(S)$ for any $S\subseteq N$,
\item {\em probabilistically representable\/} if there exists a discrete random vector $\bxi$ over $N$ such that,
for every triplet $(X,Y|Z)$ of subsets of $N$, one has
$$
X\,\ci Y\,|\,Z\,\,[h] ~\Leftrightarrow~ X\,\ci Y\,|\,Z\,\,[\bxi]\,,
$$
\item {\em weakly probabilistically representable\/} if there exists a discrete random vector $\bxi$ over $N$ such that,
for every triplet $(X,Y|Z)$ of {\em pairwise disjoint\/} subsets of $N$,
$$
X\,\ci Y\,|\,Z\,\,[h] ~\Leftrightarrow~ X\,\ci Y\,|\,Z\,\,[\bxi]\,.
$$
\end{itemize}
It is immediate that strong probabilistic representability implies the asymptotic one; by Lemma~\ref{lem.entrop-polym}, it also implies the remaining weaker versions.

A classic result by Mat\'{u}\v{s} \cite[Theorem~2]{Mat92conf} says that every {\em matroid\/} which is {\em linearly representable\/} over a {\em finite\/} field ${\dv F}$ is also strongly probabilistically representable; that particular construction of his was based on the assumption of the finiteness of the field ${\dv F}$. The same method can be used to show that this implication holds also for (integer-valued) polymatroids \cite[Theorem\,1.6.2]{Pad12}.
Note in this context that it follows from \cite[Corollary\,6.8.11]{Oxl11} that any matroid which is linearly representable over a  field is also
linearly representable over a finite field. Therefore, the assumption that the field ${\dv F}$ is finite can be omitted.

There is also a concept of (finite) {\em group representability\/}
of a polymatroid over $N$, defined in \cite[\S\,16.2]{Yeu08}, and
any polymatroid which has a group representation is strongly probabilistically representable \cite{CY02IEEE}.

To give geometric interpretation to probabilistic representability concepts
let us recall a few basic facts on the cone ${\cal H}(N)$ of rank functions $h\in {\dv R}^{{\cal P}(N)}$ of polymatroids over $N$. It is a pointed polyhedral cone and its facets are given by inequalities $\diff h(i,j|K)\geq 0$ for distinct $i,j\in N$, $K\subseteq N\setminus\{i,j\}$, and by inequalities $\diff h(i,i|N\setminus i)\geq 0$ for $i\in N$.
This means that the facets of the cone correspond to elementary triplets for representation of an augmented abstract CI structure; see Section~\ref{ssec.semi-graph}. In particular,
every {\em face\/} $F$ of ${\cal H}(N)$ can be identified with an {\em augmented abstract CI structure}, determined
by the list of elementary triplets for facets containing $F$.

The facets of ${\cal H}(N)$ specified by equalities $\diff h(i,i|N\setminus i)=0$ for $i\in N$ are quite large. In fact, there is only one extreme ray of\/ ${\cal H}(N)$ not contained in such a facet, which is the zero-one indicator
$\upsilon^{\uparrow i}$ of supersets of $\{i\}$. Let us say that a polymatroid over $N$ is {\em tight\/} if its rank function $h$ satisfies $\diff h(i,i|N\setminus i)=0$ for any $i\in N$. Thus, the set $\hat{\cal H}(N)$ of rank functions of tight polymatroids over $N$ is a face of ${\cal H}(N)$ whose extreme rays are the extreme rays of ${\cal H}(N)$ with the
exception of $|N|$ rays generated by $\upsilon^{\uparrow i}$ for $i\in N$. This is because every $h\in {\cal H}(N)$
can be assigned its tightened version $\hat{h}\in \hat{\cal H}(N)$ defined by $\hat{h}(S):= h(S) -\sum_{i\in N} \diff h(i,i|N\setminus i)\cdot \upsilon^{\uparrow i}(S)$ for $S\subseteq N$.
The facets of\/ $\hat{\cal H}(N)$ are given by inequalities $\diff h(i,j|K)\geq 0$ for distinct $i,j\in N$, $K\subseteq N\setminus\{i,j\}$ and, thus, correspond to elementary triplets for representation of a standard abstract CI structure; see Section~\ref{ssec.semi-graph}.
In particular, every {\em face\/} of\/ $\hat{\cal H}(N)$ can be identified with a {\em standard abstract CI structure}.

The first two concepts of probabilistic representability of a polymatroid with a rank function $h$
concern the {\em ray generated by $h$}. The strong representability means that the relative interior
of that ray contains an entropy function; the asymptotic representability means that it
contains a limit of entropy functions.
The probabilistic representability
concerns the {\em face of\/ ${\cal H}(N)$ generated by $h$}, which is the least face $F$ of\/ ${\cal H}(N)$ containing $h$.
It is the requirement that the relative interior of $F$ contains an entropy function.
The weak probabilistic representability concerns the {\em face $\hat{F}$ of\/ $\hat{\cal H}(N)$ generated by $\hat{h}$},
the tightened version of $h$. It is the requirement that the relative interior of the face $\hat{F}$ contains a tightened version of an entropy function.

\begin{remark}\rm\label{rem.extem-criter}
The reader may be interested in a question what are the extreme rays of the cone ${\cal H}(N)$. 
As explained above, this question is equivalent to the task what are the extreme rays of\/ $\hat{\cal H}(N)$.
It follows from the fact that ${\cal H}(N)$ is defined by inequalities with rational coefficients
that every extreme ray of\/ ${\cal H}(N)$ is generated by an integer-valued rank function.
There is a way to recognize the extremity in ${\cal H}(N)$.
Each function $h\in {\cal H}(N)$ can be identified with its {\em base polyhedron\/} $B(h)$ \cite[\S\,2.2]{Fuj91}.
Note in this context that the base polyhedron $B(h)$ is {\em not\/} the original ``polymatroid" polyhedron introduced by Edmonds \cite{Edm70}, which is named the {\em independence polyhedron\/} in \cite[\S\,2.2]{Fuj91}:
the relation is that $B(h)$ is a certain face of the latter polyhedron.
In a recent paper \cite{SK16DAM} an easy criterion was given which allows one to test
on basis of $B(h)$ whether the function $h$ generates an extreme ray of\/ ${\cal H}(N)$.
Specifically, it leads to solving a certain simple linear equation system determined by (the vertices of) $B(h)$.
Note, however, that the paper \cite{SK16DAM} is primarily devoted to supermodular functions, which are mirror images of submodular functions. The point is that the cone $\hat{\cal H}(N)$ of rank functions for tight polymatroids is isomorphic to the cone of (suitably) standardized supermodular functions. The class of base polyhedra for polymatroids then coincides with the class of polytopes that are known as ``cores" of supermodular (= convex) games. The same class of polytopes is also known as the class of  {\em generalized permutohedra\/} \cite{PRW08}.
\end{remark}

\subsection{CI structures over four discrete variables}\label{ssec.CI4}
Here we describe in more detail the methodological approach from \cite{MatStu95CPC,Mat95CPC,Mat99CPC}.
The basic idea was the embedding $\bxi\mapsto h_{\ibxi}$ of discrete random vectors over $N$ into the cone ${\cal H}(N)$; see Section~\ref{ssec.entropy}.

The starting conjecture (of two of us), which holds in case $|N|=3$, was that every $h\in {\cal H}(N)$ is a multiple of some entropy function. To verify that it would be enough, thanks to the construction in Section~\ref{ssec.lattice},
to find, for every extreme ray $R$ of\/ ${\cal H}(N)$, a discrete random vector $\bxi$ over $N$ with $h_{\ibxi}$ generating $R$.
Mat\'{u}\v{s} found the extreme rays of\/ ${\cal H}(N)$ in case $|N|=4$ in a previous paper of his \cite[\S\,5]{Mat94DM}:
there are 41 of them and they break into 11 permutational types.

We found the desired probabilistic constructions for 10 permutational types \cite[\S\,5]{MatStu95CPC}
and were stuck with the last permutational type containing 6 rays. Note in this context
that all these 35 rays, for which we found the constructions, are generated by integer-valued rank functions
linearly representable over finite fields. Thus, the existence of their (strong) probabilistic representations could also be derived from results
on linearly representable polymatroids \cite{Mat92conf,Pad12} mentioned in Section~\ref{ssec.repre-polym}.
In the Appendix, Section~\ref{sec.sub-max-constr}, we give 9 of these constructions, namely those for tight polymatroids.

The breakthrough was an observation by Mat\'{u}\v{s} that the last permutational type is not
strongly probabilistically representable \cite[\S\,7]{Mat94IJGS}. The motivational source
for that observation was 
his perception that the rank function
$$
h_{xy}(S) ~:=~
\left\{
\begin{array}{cl}
0 & \mbox{for $S=\emptyset$,}\\
4 & \mbox{for $S=xy$ and $S=N$,}\\
|S|+1 & \mbox{for other $S\subseteq N$,}
\end{array}
\right.
~ \mbox{where $\{x,y\}\subset N$, $|N|=4$,}
$$
generating a ray of the last type, yields a {\em linearly non-representable} polymatroid. The argument
for the latter claim was that $h_{xy}$ can be obtained by coarsening from the rank function
of the simplest {\em non-linear matroid}, the so-called {\em V\'{a}mos cube}, defined on an 8-element basic set.
The observation from \cite[Lemma\,1]{Mat94IJGS} that $h_{xy}$ has no strong probabilistic representation was based on a derived CI property which involves functional dependence;
therefore, it does not exclude the (pertinent) existence of a weak probabilistic representation for $h_{xy}$.
The conclusion that $h_{xy}$ has no weak probabilistic representation, however,
follows from standard CI implications found in \cite[\S\,5]{Stu94IJGS}
on basis of \cite[Lemma\,1]{Mat94IJGS}.
\smallskip

A supportive argument for linear non-representability of $h_{xy}$ is that it
does not satisfy the {\em Ingleton inequality\/} $\ingl h_{xy}(x,y)\geq 0$. This
special inequality was published in 1971: Ingleton \cite{Ing71conf} showed that the rank function
of any linearly representable {\em matroid\/} over $N$, $|N|=4$, must comply with it.
An interesting geometric observation \cite[Lemma\,4.1]{MatStu95CPC}
is that the sub-cone of\/ ${\cal H}(N)$ demarcated by adding six versions of Ingleton inequality
has 35 extreme rays, namely those rays of\/ ${\cal H}(N)$ for which the probabilistic representation was found.
This lead us to a revised conjecture that the sub-cone defined by adding Ingleton inequalities to
those for ${\cal H}(N)$ determines all probabilistic CI structures over $N$, that is, probabilistic and linear representability coincide in case of 4 variables.
\smallskip

The revised conjecture appeared not to be true: we found four examples
of probability distributions whose induced entropy functions do not satisfy the Ingleton
inequality; see Examples \ref{exa.M-1}-\ref{exa.M-4} in Section~\ref{ssec.counter-Ingleton}.
Thus, we got examples of probabilistically representable polymatroids that
are not linearly representable.
Mat\'{u}\v{s} then continued in his laborious 
analysis of numerous remaining cases of ``un-decided"
abstract CI structures. He obtained them on basis of a more detailed geometric analysis of the cone ${\cal H}(N)$
from \cite[\S\,6]{MatStu95CPC}. He found many additional CI implications in the next paper \cite{Mat95CPC}
by combining various methods 
but stuck in 1995 with a couple of undecided cases.
A breakthrough came in 1997 when Zhang and Yeung \cite{ZhaYeu97ITI} published their
first conditional information inequality. Mat\'{u}\v{s} found out that he is able to
settle one of his remaining cases by deriving a CI implication as a consequence of
their inequality. Moreover, in \cite[Proposition\,2.1]{Mat99CPC} he proved the second conditional information inequality and solved the last case by deriving
the last CI implication. 

\subsection{Information inequalities}\label{ssec.infor-ineq}
In 1998 Zhang and Yeung \cite{ZhaYeu98ITI} found the first unconstrained linear inequality
for the entropy function which is not implied by the inequalities defining ${\cal H}(N)$ and
showed that the set of limits of entropy functions is closed convex cone.
Thus, they gave
the first negative response to a general question raised by Pippenger \cite{Pip86} in 1986: what are the
``laws of information theory", specifically whether classic Shannon inequalities
characterize entropy functions. In 2007 then Mat\'{u}\v{s} \cite{Mat07ISIT} showed that the
cone of limits of entropy functions is not polyhedral if $|N|\geq 4$, which means there is an infinite number
of unconstrained information inequalities (= linear inequalities valid for entropy functions).
\smallskip

The interest in conditional (= constrained) information inequalities has been restored in the 2010s
when Kaced and Romashchenko \cite{KacRom13IEEE} started to study these inequalities from a theoretical point of view.
Besides the proofs of several new inequalities they proposed to classify these inequalities.
Some of the conditional information inequalities can be derived from unconstrained ones, which means
they are essentially unconstrained. Kaced and Romashchenko, however, showed that
two former conditional information inequalities from \cite{ZhaYeu97ITI,Mat99CPC} and their own
new inequalities are {\em essentially conditional}, which means
they cannot be derived from unconstrained linear information inequalities.

They also introduced more detailed classification of essentially conditional inequalities:
some of them hold solely for entropy functions and some of them hold even for limits of
entropy functions. By a special evidence technique they showed that most of (their) new
inequalities are valid for limits of entropy functions. On the other hand, they
gave examples that two discussed essentially conditional inequalities need not hold
for limits of entropy functions; the first conditional information inequality \cite{ZhaYeu97ITI}
is one of them.

\section{Conditional Ingleton inequalities}\label{sec.condi-Ingleton}
In this section we present sufficient conditions in terms of CI 
for the validity of Ingleton inequality, give a few counter-examples to its general probabilistic
validity and raise related open questions.

\subsection{Sufficient conditions}\label{ssec.suff-Ingleton}
Our main theoretical result offers five conditions ensuring the validity of the Ingleton inequality;
to fix the context we consider the case $\ingl (X,Y)\geq 0$.

\begin{theorem}\label{thm.main}
Given the entropy function $h$ induced by a discrete random vector over $N$, the following
implications hold:
\begin{description}
\item[\mbox{[\,1cI\,]}]~~ $\diff h(X,Y|\emptyset)=0=\diff h(X,Y|Z)\quad \Rightarrow\quad \ingl h(X,Y)\geq 0$,
\item[\mbox{[\,2cI\,]}]~~ $\diff h(X,Y|Z)=0=\diff h(Y,U|Z)\quad \Rightarrow\quad \ingl h(X,Y)\geq 0$,
\item[\mbox{[\,3cI\,]}]~~ $\diff h(X,Z|U)=0=\diff h(X,U|Z)\quad \Rightarrow\quad \ingl h(X,Y)\geq 0$,
\item[\mbox{[\,4cI\,]}]~~ $\diff h(X,Z|U)=0=\diff h(Z,U|X)\quad \Rightarrow\quad \ingl h(X,Y)\geq 0$,
\item[\mbox{[\,5cI\,]}]~~ $\diff h(X,Z|U)=0=\diff h(Y,Z|U)\quad \Rightarrow\quad \ingl h(X,Y)\geq 0$.
\end{description}
\end{theorem}

\begin{proof}
This shown in the Appendix, Section~\ref{sec.appendix-proofs}.
\end{proof}

Because of the symmetry of the expression $\ingl h(X,Y)$ with respect to $X\leftrightarrow Y$ and $Z\leftrightarrow U$
the conditional Ingleton inequalities from Theorem~\ref{thm.main} have altogether 14 possible forms;
the inequalities [\,2cI\,] and [\,4cI\,] have four possible permutated versions while [\,1cI\,], [\,3cI\,], and [\,5cI\,] only two of them.

\subsection{Counter-examples}\label{ssec.counter-Ingleton}
To show that the premises for Ingleton inequality $\ingl h(X,Y)\geq 0$ in Theorem~\ref{thm.main} are minimal \mbox{possible} ones
one needs counter-examples to its general validity. Four of them are based on constructions
of probability distributions having certain special CI structure;
analogous constructions were formerly given in \cite{Mat95CPC,Mat99CPC,StuBoc94}.
\medskip

The first counter-example is loosely related to (M.1) form of\/ $\ingl (X,Y)$.

\begin{example}\rm\label{exa.M-1}
There exists a discrete random vector (over $N$) such that
$$
[~ X\ci Y\,|\,\emptyset ~~\&~~ Z\ci U\,|\,X ~~\&~~ Z\ci U\,|\,Y ~~\&~~ Z\ci U\,|\,XY ~] ~~\not\Rightarrow~~
\ingl (X,Y)\geq 0\,.
$$
Put $N=\{x,y,z,u\}$ and define the density $p$ of a binary random vector:
\begin{center}
\begin{tabular}{lclc}
& $x~y~z~u$ &&\\
$p:$ & $(\0 ,\0 ,\0 ,\0 )$ &$\mapsto$ & $1/4$ \\
& $(\0 ,\1 ,\0 ,\1 )$ &$\mapsto$ & $1/4$ \\
& $(\1 ,\0 ,\0 ,\1 )$ &$\mapsto$ & $1/4$ \\
& $(\1 ,\1 ,\1 ,\1 )$ &$\mapsto$ & $1/4$
\end{tabular}
\quad with zero values for other configurations.
\end{center}
The reader can easily check the validity of above CI statements.
Using (M.1) in Lemma~\ref{lem.Inglet-forms}
one has $\ingl (X,Y)= -\diff (Z,U|\emptyset)<0$ because $Z\notci U\,|\,\emptyset$.
The induced CI structure over $N$ 
has 6 permutably equivalent versions.
\end{example}

The second counter-example is based on (M.2) form of\/ $\ingl (X,Y)$.

\begin{example}\rm\label{exa.M-2}
There exists a discrete random vector such that
$$
[~ X\ci Y\,|\,\emptyset ~~\&~~ X\ci Z\,|\,U ~~\&~~ Z\ci U\,|\,Y ~~\&~~ Z\ci U\,|\,XY ~] ~~\not\Rightarrow~~
\ingl (X,Y)\geq 0\,.
$$
To this end we put $N=\{x,y,z,u\}$ and define the density $p$ of a binary random vector over $N$ as follows:
\begin{center}
\begin{tabular}{lclc}
& $x~y~z~u$ &&\\
$p:$
& $(\0 ,\0 ,\0 ,\0 )$ &$\mapsto$ & $1/4$ \\
& $(\0 ,\1 ,\1 ,\0 )$ &$\mapsto$ & $1/4$ \\
& $(\1 ,\0 ,\0 ,\1 )$ &$\mapsto$ & $1/4$ \\
& $(\1 ,\1 ,\0 ,\0 )$ &$\mapsto$ & $1/8$ \\
& $(\1 ,\1 ,\1 ,\0 )$ &$\mapsto$ & $1/8$
\end{tabular}
\quad with zero values for other configurations.
\end{center}
The reader can check directly the validity of above CI statements.
Using (M.2) in Lemma~\ref{lem.Inglet-forms}
one has $\ingl (X,Y)= -\diff (X,Z|\emptyset)<0$ because $X\notci Z\,|\,\emptyset$.
The induced CI structure over $N$ 
has 24 permutably equivalent versions.
\end{example}

The third counter-example is related to (M.3) and (M.5) forms of\/ $\ingl (X,Y)$.

\begin{example}\rm\label{exa.M-3-5}
There exists a discrete random vector such that
$$
[~ X\ci Y\,|\,Z ~~\&~~ X\ci Z\,|\,U ~~\&~~ Z\ci U\,|\,Y ~~\&~~ Z\ci U\,|\,XY ~] ~~\not\Rightarrow~~
\ingl (X,Y)\geq 0\,.
$$
We put $N=\{x,y,z,u\}$ and define the density $p$ of a binary random vector:
\begin{center}
\begin{tabular}{lclc}
& $x~y~z~u$ &&\\
$p:$ & $(\0 ,\0 ,\0 ,\1 )$ &$\mapsto$ & $2/7$ \\
& $(\0 ,\0 ,\1 ,\1 )$ &$\mapsto$ & $1/7$ \\
& $(\0 ,\1 ,\1 ,\1 )$ &$\mapsto$ & $1/7$ \\
& $(\1 ,\0 ,\0 ,\1 )$ &$\mapsto$ & $1/7$ \\
& $(\1 ,\0 ,\1 ,\1 )$ &$\mapsto$ & $1/7$ \\
& $(\1 ,\1 ,\1 ,\0 )$ &$\mapsto$ & $1/7$
\end{tabular}
\quad with zero values for other configurations.
\end{center}
The reader can directly verify the validity of above CI statements.
Using (M.3) in Lemma~\ref{lem.Inglet-forms}
one has $\ingl (X,Y)= -\diff (X,Z|Y)<0$ as $X\notci Z\,|\,Y$;  by (M.5),
alternatively, one has $\ingl (X,Y)= -\diff (X,Z|YU)<0$ since $X\notci Z\,|\,YU$.
The induced CI structure over $N$ 
has 24 permutably equivalent versions.
\end{example}

The fourth counter-example is loosely related to (M.4) form of\/ $\ingl (X,Y)$.

\begin{example}\rm\label{exa.M-4}
There exists a discrete random vector such that
\begin{eqnarray*}
\lefteqn{\hspace*{-42mm}[~ X\ci Y\,|\,Z ~~\&~~   X\ci Y\,|\,U ~~\&~~  Z\ci U\,|\,X ~~\&~~  Z\ci U\,|\,Y ~~\&~~ Z\ci U\,|\,XY ~]}\\
 &\not\Rightarrow& \ingl (X,Y)\geq 0\,.
\end{eqnarray*}
To this end we put $N=\{x,y,z,u\}$ and define the density $p$ of a binary random vector over $N$ as follows:
\begin{center}
\begin{tabular}{lclc}
& $x~y~z~u$ &&\\
$p:$ & $(\0 ,\0 ,\0 ,\0 )$ &$\mapsto$ & $2/7$ \\
& $(\0 ,\1 ,\0 ,\0 )$ &$\mapsto$ & $1/7$ \\
& $(\0 ,\1 ,\1 ,\0 )$ &$\mapsto$ & $1/7$ \\
& $(\1 ,\0 ,\0 ,\0 )$ &$\mapsto$ & $1/7$ \\
& $(\1 ,\0 ,\0 ,\1 )$ &$\mapsto$ & $1/7$ \\
& $(\1 ,\1 ,\0 ,\0 )$ &$\mapsto$ & $1/7$ \\
\end{tabular}
\quad with zero values for other configurations.
\end{center}
The reader can directly verify the validity of above CI statements.
Using (M.4) in Lemma~\ref{lem.Inglet-forms}
one has $\ingl (X,Y)= -\diff (X,Y|ZU)<0$ because $X\notci Y\,|\,ZU$.
The induced CI structure over $N$ 
has 6 permutably equivalent versions.
\end{example}

The preceding examples allow one to observe the following.

\begin{cor}\rm\label{cor.minimal}
The premises of the implications from Theorem~\ref{thm.main} are minimal.
\end{cor}

\begin{proof}
No single premise $\diff h (A,B|C)=0$ there implies\/ $\ingl h(X,Y)\geq 0$.
\end{proof}

An important addition to Theorem~\ref{thm.main} is the following (see later Remark~\ref{rem.approach}).

\begin{example}\rm\label{exa.sixth}
There exists a discrete random vector (over $N$) such that
$$
[~ X\ci Z\,|\,U ~~\&~~ Y\ci U\,|\,Z  ~] ~~\not\Rightarrow~~ \ingl (X,Y)\geq 0\,.
$$
To show that we put $N=\{x,y,z,u\}$ and define the density $p$ of a binary random vector over $N$ as follows:
\begin{center}
\begin{tabular}{lclc}
& $x~y~z~u$ &&\\
$p:$ & $(\0 ,\0 ,\0 ,\0 )$ &$\mapsto$ & $18/64$ \\[0.4ex]
& $(\0 ,\1 ,\0 ,\0 )$ &$\mapsto$ & $~3/64$ \\
& $(\0 ,\1 ,\0 ,\1 )$ &$\mapsto$ & $~1/64$ \\
& $(\0 ,\1 ,\1 ,\0 )$ &$\mapsto$ & $~7/64$ \\
& $(\0 ,\1 ,\1 ,\1 )$ &$\mapsto$ & $~3/64$ \\[0.4ex]
& $(\1 ,\0 ,\0 ,\0 )$ &$\mapsto$ & $~3/64$ \\
& $(\1 ,\0 ,\0 ,\1 )$ &$\mapsto$ & $~7/64$ \\
& $(\1 ,\0 ,\1 ,\0 )$ &$\mapsto$ & $~1/64$ \\
& $(\1 ,\0 ,\1 ,\1 )$ &$\mapsto$ & $~3/64$ \\[0.4ex]
& $(\1 ,\1 ,\1 ,\1 )$ &$\mapsto$ & $18/64$
\end{tabular}
\quad with zero values for other configurations.
\end{center}
It is easy to compute the respective marginal densities for $XZU$ and $YZU$
\begin{center}
\begin{tabular}{clc}
$x~z~u$ &&\\
$(\0 ,\0 ,\0)$ &$\mapsto$ & $21/64$ \\
$(\0 ,\0 ,\1)$ &$\mapsto$ & $~1/64$ \\
$(\0 ,\1 ,\0)$ &$\mapsto$ & $~7/64$ \\
$(\0 ,\1 ,\1)$ &$\mapsto$ & $~3/64$ \\[0.4ex]
$(\1 ,\0 ,\0)$ &$\mapsto$ & $~3/64$ \\
$(\1 ,\0 ,\1)$ &$\mapsto$ & $~7/64$ \\
$(\1 ,\1 ,\0)$ &$\mapsto$ & $~1/64$ \\
$(\1 ,\1 ,\1)$ &$\mapsto$ & $21/64$
\end{tabular}
\qquad
\begin{tabular}{clc}
$y~z~u$ &&\\
$(\0 ,\0 ,\0)$ &$\mapsto$ & $21/64$ \\
$(\0 ,\0 ,\1)$ &$\mapsto$ & $~7/64$ \\
$(\0 ,\1 ,\0)$ &$\mapsto$ & $~1/64$ \\
$(\0 ,\1 ,\1)$ &$\mapsto$ & $~3/64$ \\[0.4ex]
$(\1 ,\0 ,\0)$ &$\mapsto$ & $~3/64$ \\
$(\1 ,\0 ,\1)$ &$\mapsto$ & $~1/64$ \\
$(\1 ,\1 ,\0)$ &$\mapsto$ & $~7/64$ \\
$(\1 ,\1 ,\1)$ &$\mapsto$ & $21/64$
\end{tabular}
\end{center}
and observe that $X\ci Z\,|\,U$ and $Y\ci U\,|\,Z$ (using the cross-product criterion).
The goal is to show that the induced entropy function $h$ satisfies $\ingl h(X,Y)< 0$.
Note that two-dimensional marginals have a special form $P_{\alpha}$ with a density
\begin{center}
\begin{tabular}{lclc}
$p_{\alpha}:$ & $(\0 ,\0)$ &$\mapsto$ & $\frac{1}{4}\cdot (1+\alpha)$ \\
&$(\0 ,\1)$ &$\mapsto$ & $\frac{1}{4}\cdot (1-\alpha)$ \\
&$(\1 ,\0)$ &$\mapsto$ & $\frac{1}{4}\cdot (1-\alpha)$ \\
&$(\1 ,\1)$ &$\mapsto$ & $\frac{1}{4}\cdot (1+\alpha)$
\end{tabular}
\qquad
for some $\alpha\in [0,1]$.
\end{center}
Specifically, one has $\alpha_{xy}=1/8$, $\alpha_{xz}=\alpha_{yu}=3/8$ and $\alpha_{zu}=1/2$.
Observe that the mutual information for the distribution $P_{\alpha}$ is
$$
I(\alpha) ~:=~ \frac{1}{2}\cdot \left[\, (1+\alpha)\cdot \ln (1+\alpha) ~+~ (1-\alpha)\cdot \ln (1-\alpha)\,
\right] \,.
$$
In particular, $\diff h(X,Y|\emptyset)=I(1/8)$, $\diff h(X,Z|\emptyset)=\diff h(Y,U|\emptyset)=I(3/8)$ and
$\diff h(Z,U|\emptyset)=I(1/2)$. This allows one to write the $(16)$-multiple of the Ingleton expression in the
following form:
\begin{eqnarray*}
\lefteqn{16\cdot\ingl h(X,Y) \stackrel{\eqref{eq.Inglet}}{=} \underbrace{16\cdot\diff h (X,Z|U)}_{=0} +  \underbrace{16\cdot\diff h (Y,U|Z)}_{=0}}\\
&& +16\cdot\diff h (X,Y|\emptyset) +  16\cdot\diff h (Z,U|\emptyset) - 16\cdot\diff h (X,Z|\emptyset) - 16\cdot\diff h (Y,U|\emptyset)\\
&=& 16\cdot [\,I(1/8) + I(1/2) - 2\cdot I(3/8)\,]\\
&=& 9\cdot \ln \frac{9}{8} +7\cdot \ln \frac{7}{8} + 12\cdot \ln \frac{3}{2} +4\cdot \ln \frac{1}{2}
-22\cdot \ln \frac{11}{8} -10\cdot \ln \frac{5}{8}\\
&=& 32\cdot \ln (2) + 30\cdot \ln (3) -10\cdot \ln (5) + 7\cdot \ln (7) - 22\cdot \ln (11)\\
&\doteq& -0.0876256 ~<~  0\,.
\end{eqnarray*}
Hence, $\ingl h(X,Y)<0$, which concludes the arguments. Note that the induced CI structure
involves only 2 non-trivial statements
$X\ci Z\,|\,U$ and $Y\ci U\,|\,Z$.
\end{example}

\subsection{Remaining open tasks}\label{ssec.open-Ingleton}
First, let us recall the considerations leading to Theorem~\ref{thm.main}.

\begin{remark}\rm\label{rem.approach}
The simple question at the beginning was which CI assumptions \mbox{imply} the validity of
Ingleton inequality\, $\ingl (X,Y)\geq 0$. Clearly, by Lemma~\ref{lem.Inglet-forms},
there are 14 single CI statements implying its validity, namely the ones whose respective terms
are with ``minus" in one of the 14 basic forms of the Ingleton expression.
Thus, other possible CI assumptions for\, $\ingl (X,Y)\geq 0$ are subsets of the 10-element
set of remaining CI statements composed of $X,Y,Z,U$.

The constructions from Examples~\ref{exa.M-1}-\ref{exa.M-4} were known formerly \cite{Mat95CPC,Mat99CPC,StuBoc94};
these appear to be maximal CI assumptions which {\em do not imply\/}\, \mbox{$\ingl (X,Y)\geq 0$.}
This led to the task to determine the minimal subsets of the above 10-element set of CI statements
that are {\em not contained\/} in one of those non-implicators. These are 2-element sets falling
into 6 permutational types (relative to \mbox{$X\leftrightarrow Y$} and $Z\leftrightarrow U$).
Five of them, reported in Theorem~\ref{thm.main}, appeared to imply\/ \mbox{$\ingl (X,Y)\geq 0$,}
while the sixth type not, as shown by Example~\ref{exa.sixth}.
\end{remark}

Thus, Example~\ref{exa.sixth} can be viewed as a counter-example
to probabilistic validity of the sixth potential conditional Ingleton inequality.
Nevertheless, it does not exclude its existence. Therefore, we ask the following.

\begin{open}\rm
Is there a set ${\cal X}$ of CI statements such that
$$
\{\, X\ci Z\,|\,U,\, Y\ci U\,|\,Z \,\} \subset {\cal X} \subseteq
\{\, X\ci Y\,|\,\emptyset,\, X\ci Z\,|\,U,\, Y\ci U\,|\,Z,\, Z\ci U\,|\,XY \,\}
$$
which implies the validity of\, $\ingl (X,Y)\geq 0$\,?
\end{open}

Note in this context that it follows from 
Theorem~\ref{thm.main} and Example~\ref{exa.sixth} that these are all the remaining potential CI implicators for\/ $\ingl (X,Y)\geq 0$.
Another couple of open problems is inspired by concepts and results from \cite{KacRom13IEEE}.

\begin{open}\rm
Is the new inequality [\,5cI\,] from Theorem~\ref{thm.main}
essentially conditional? 
Which of the inequalities [\,1cI\,]-[\,5cI\,] are valid for functions 
that are limits of entropy functions induced by discrete random vectors? 
\end{open}

Remark that, by results from \cite{KacRom13IEEE}, the inequalities [\,1cI\,]-[\,4cI\,] are essentially conditional
and [\,3cI\,]-[\,4cI\,] hold for limits of entropy functions.

\section{Conditional independence structures revisited}\label{sec.CI-structures}
In this section we apply the previously presented results to characterize (abstract properties of)
CI structures induced by four discrete random variables.

\subsection{CI equivalences}\label{ssec.CI-equiv}
There are properties of (discrete) probabilistic CI structures
which follow solely from the basic fact that every entropy function is a polymatroid (see Lemma~\ref{lem.entrop-polym}).
These properties can be viewed as two-way CI implications where one set of CI statements implies
another set of CI statements and conversely.

\begin{cor}\em\label{cor.CI-equiv}
Every (standard) abstract CI structure over $N$ induced by the rank function of a 
polymatroid  is a semi-graphoid which complies with the following five conditions:
given pairwise disjoint sets $X,Y,Z,U\subseteq N$, one has
\begin{itemize}
\item[(E:1)] $[~ X\ci Y\,|\,Z ~~~\&~~~ X\ci Z\,|\,U  ~~~\&~~~ X\ci U\,|\,Y ~]$\\
 ~~~$\Leftrightarrow~ [~ X\ci Y\,|\,U ~~~\&~~~ X\ci Z\,|\,Y ~~~\&~~~ X\ci U\,|\,Z  ~]$,
\item[(E:2)] $[~ X\ci Y\,|\,Z ~~~\&~~~ X\ci U\,|\,Y ~~~\&~~~ Y\ci Z\,|\,U  ~~~\&~~~ Z\ci U\,|\,X  ~]$\\
 ~~~$\Leftrightarrow~ [~ X\ci Y\,|\,U ~~~\&~~~ X\ci U\,|\,Z ~~~\&~~~ Y\ci Z\,|\,X  ~~~\&~~~ Z\ci U\,|\,Y  ~]$,
\item[(E:3)] $[~ X\ci Y\,|\,ZU ~~~\&~~~ X\ci Z\,|\,\emptyset ~~~\&~~~  Y\ci U\,|\,\emptyset ~~~\&~~~ Z\ci U\,|\,XY  ~]$\\
 ~~~$\Leftrightarrow~ [~  X\ci Y\,|\,\emptyset ~~~\&~~~ X\ci Z\,|\,YU ~~~\&~~~ Y\ci U\,|\,XZ  ~~~\&~~~Z\ci U\,|\,\emptyset ~]$,
\item[(E:4)] $[~ X\ci Y\,|\,\emptyset  ~~~\&~~~ X\ci Y\,|\,ZU ~~~\&~~~ Z\ci U\,|\,X  ~~~\&~~~ Z\ci U\,|\,Y  ~]$\\
 ~~~$\Leftrightarrow~ [~ X\ci Y\,|\,Z  ~~~\&~~~ X\ci Y\,|\,U ~~~\&~~~ Z\ci U\,|\,\emptyset ~~~\&~~~ Z\ci U\,|\,XY  ~]$,
\item[(E:5)] $[~ X\ci Y\,|\,ZU ~~~\&~~~ X\ci U\,|\,Y  ~~~\&~~~ Y\ci Z\,|\,\emptyset ~~~\&~~~ Z\ci U\,|\,X  ~]$\\
 ~~~$\Leftrightarrow~ [~  X\ci Y\,|\,U  ~~~\&~~~ X\ci U\,|\,YZ ~~~\&~~~ Y\ci Z\,|\,X ~~~\&~~~ Z\ci U\,|\,\emptyset ~]$.
\end{itemize}
In particular, every (standard) probabilistic CI structure satisfies (E:1)-(E:5).
\end{cor}

\begin{proof}
Lemma~\ref{lem.polym-semig} says that the induced CI structure is a semi-graphoid.
To verify the implication $\Rightarrow$ in (E:1), write for a rank function $r:{\cal P}(N)\to {\dv R}$:
\begin{eqnarray*}
\lefteqn{\hspace*{-8mm}0= \diff r(X,Y|Z) + \diff r(X,Z|U)  + \diff r(X,U|Y)}\\
&=& +\, r(XY)+r(XZ)+r(XU)+r(YZ)+r(YU)+r(ZU)\\
&& -\, r(Y)-r(Z)-r(U)-r(XYZ)-r(XYU)-r(XZU)\\
&=& \diff r (X,Y|U) + \diff r (X,Z|Y)  + \diff r (X,U|Z)
\end{eqnarray*}
and deduce from non-negativity of the involved difference terms that they have to vanish.
The opposite implication $\Leftarrow$ in (E:1) and also all other implications in (E:2)-(E:5) can be verified
in an analogous way; this is left to the reader. The last claim then easily follows from
the equivalence \eqref{eq.CI-entrop} in Lemma~\ref{lem.entrop-polym}.
\end{proof}

The CI equivalences from Corollary~\ref{cor.CI-equiv}, including the properties defining a semi-graphoid,
appear to be all CI equivalences which effectively apply in case four random variables.
Note, however, that one can derive many further valid CI equivalences in the same way 
in case five and more discrete variables; one can even use computers for this purpose \cite{BHLS10}.

\begin{remark}\em\label{rem.structural}
This is to commemorate the occurrence of the properties from Corollary~\ref{cor.CI-equiv} in the literature.
The condition (E:4) appeared as (A.3) in \cite{Stu89} as the first ever CI property
which does not follow from
semi-graphoid axioms. The conditions (E:1)-(E:5) were all formulated in \cite{Stu94IJGS}
as properties (A.3)-(A.7) of the so-called ``structural" semi-graphoids discussed there;
note that one can show using \cite[\S\,5.3]{Stu05} that ``structural" semi-graphoids
coincide with standard CI structures induced by polymatroids.
On the other hand, the conditions (E:1)-(E:5) were not mentioned
in \cite{MatStu95CPC,Mat95CPC,Mat99CPC}; they are, however, implicit 
in the concept of a ``semi-matroid" defined there, which concept corresponds to
an augmented CI structure induced by a polymatroid.
\end{remark}

\subsection{CI implications}
Here we apply the conditional Ingleton inequalities from Theorem~\ref{thm.main}
to derive CI implications valid for discrete random vectors.

\begin{cor}\em\label{cor.CI-implic}
Every (standard) probabilistic CI structure induced by a discrete random vector over $N$
complies with the following 19 conditions:
given pairwise disjoint sets $X,Y,Z,U\subseteq N$, one has
\begin{itemize}
\item[(I:1)] $[\, X\ci Y\,|\,\emptyset ~\,\&~\, X\ci Y\,|\,Z  ~\,\&~\, Z\ci U\,|\,X ~\,\&~\, Z\ci U\,|\,Y ~]
 ~~\Rightarrow~ Z\ci U\,|\,\emptyset$,
\item[(I:2)] $[\, X\ci Y\,|\,\emptyset ~\,\&~\, X\ci Z\,|\,U  ~\,\&~\, Z\ci U\,|\,X ~\,\&~\, Z\ci U\,|\,Y ~]
 ~~\Rightarrow~ Z\ci XU\,|\,\emptyset$,
\item[(I:3)] $[\, X\ci Y\,|\,\emptyset ~\,\&~\, X\ci Y\,|\,U  ~\,\&~\, X\ci Z\,|\,U ~\,\&~\, Z\ci U\,|\,Y ~]
 ~~\Rightarrow~ X\ci Z\,|\,\emptyset$,
\item[(I:4)] $[\, X\ci Y\,|\,\emptyset ~\,\&~\, X\ci Z\,|\,U  ~\,\&~\, X\ci U\,|\,Z ~\,\&~\, Z\ci U\,|\,Y ~]
 ~~\Rightarrow~ X\ci ZU\,|\,\emptyset$,
\item[(I:5)] $[\, X\ci Y\,|\,\emptyset ~\,\&~\, X\ci Z\,|\,U  ~\,\&~\, Y\ci U\,|\,Z ~\,\&~\, Z\ci U\,|\,Y ~]
 ~~\Rightarrow~ X\ci Z\,|\,\emptyset$,
\item[(I:6)] $[\, X\ci Y\,|\,\emptyset ~\,\&~\, X\ci Z\,|\,U  ~\,\&~\, Y\ci Z\,|\,U ~\,\&~\, Z\ci U\,|\,Y ~]
 ~~\Rightarrow~ X\ci Z\,|\,\emptyset$,
\item[(I:7)] $[\, X\ci Y\,|\,\emptyset ~\,\&~\, X\ci Y\,|\,Z  ~\,\&~\, X\ci Z\,|\,U ~\,\&~\, Z\ci U\,|\,Y ~]
 ~~\Rightarrow~ X\ci YZ\,|\,\emptyset$,
\item[(I:8)] $[\, X\ci Y\,|\,Z ~\,\&~\, X\ci Z\,|\,U  ~\,\&~\, Y\ci U\,|\,Z ~\,\&~\, Z\ci U\,|\,Y ~]
 ~~\Rightarrow~ X\ci Z\,|\,Y$,
\item[(I:9)] $[\, X\ci Y\,|\,Z ~\,\&~\, X\ci Y\,|\,U  ~\,\&~\, X\ci Z\,|\,U ~\,\&~\, Z\ci U\,|\,Y ~]
 ~~\Rightarrow~ X\ci Z\,|\,Y$,
\item[(I:10)] $[\, X\ci Y\,|\,Z ~\,\&~\, X\ci Z\,|\,U  ~\,\&~\, X\ci U\,|\,Z ~\,\&~\, Z\ci U\,|\,Y ~]
 ~~\Rightarrow~ X\ci Z\,|\,Y$,
\item[(I:11)] $[\, X\ci Y\,|\,Z ~\,\&~\, X\ci Z\,|\,U  ~\,\&~\, Z\ci U\,|\,X ~\,\&~\, Z\ci U\,|\,Y ~]
 ~~\Rightarrow~ X\ci Z\,|\,Y$,
\item[(I:12)] $[\, X\ci Y\,|\,Z ~\,\&~\, X\ci Z\,|\,U  ~\,\&~\, Y\ci Z\,|\,U ~\,\&~\, Z\ci U\,|\,Y ~]
 ~~\Rightarrow~ X\ci Z\,|\,Y$,
\item[(I:13)] $[\, X\ci Y\,|\,\emptyset ~\,\&~\, X\ci Y\,|\,Z ~\,\&~\, X\ci Y\,|\,U ~\,\&~\, Z\ci U\,|\,XY ~]
 ~\Rightarrow~ X\ci Y\,|\,ZU$,
\item[(I:14)] $[\, X\ci Y\,|\,Z ~\,\&~\, X\ci Y\,|\,U ~\,\&~\, X\ci Z\,|\,U ~\,\&~\, Z\ci U\,|\,XY ~]
 ~\Rightarrow~ X\ci YZ\,|\,U$,
\item[(I:15)] $[\, X\ci Y\,|\,\emptyset ~\,\&~\, X\ci Y\,|\,Z ~\,\&~\, X\ci Z\,|\,U ~\,\&~\, Z\ci U\,|\,XY ~]
 ~\Rightarrow~ X\ci Z\,|\,YU$,
\item[(I:16)] $[\, X\ci Y\,|\,Z ~\,\&~\, X\ci Z\,|\,U  ~\,\&~\, Y\ci U\,|\,Z ~\,\&~\, Z\ci U\,|\,XY ~]
 ~\Rightarrow~ X\ci Z\,|\,YU$,
\item[(I:17)] $[\, X\ci Y\,|\,Z ~\,\&~\, X\ci Z\,|\,U  ~\,\&~\, X\ci U\,|\,Z ~\,\&~\, Z\ci U\,|\,XY ~]
 ~\Rightarrow~ X\ci Z\,|\,YU$,
\item[(I:18)] $[\, X\ci Y\,|\,Z ~\,\&~\, X\ci Z\,|\,U  ~\,\&~\, Z\ci U\,|\,X ~\,\&~\, Z\ci U\,|\,XY ~]
 ~\Rightarrow~ X\ci Z\,|\,YU$,
\item[(I:19)] $[\, X\ci Y\,|\,Z ~\,\&~\, X\ci Z\,|\,U  ~\,\&~\, Y\ci Z\,|\,U ~\,\&~\, Z\ci U\,|\,XY ~]
 ~\Rightarrow~ Z\ci XY\,|\,U$.
\end{itemize}
\end{cor}
\medskip


\begin{proof}
In this proof, we will use, besides five conditional Ingleton inequalities from Theorem~\ref{thm.main}
in their basic forms [\,1cI\,], [\,2cI\,], \ldots , [\,5cI\,] written there, also alternative versions of two of them, namely
\begin{description}
\item[\mbox{[\,cI2\,]}]~~  $\diff h(X,Y|U)=0=\,\diff h(X,Z|U)\quad \Rightarrow\quad \ingl h(X,Y)\geq 0$,
\item[\mbox{[\,cI4\,]}]~~  $\diff h(Y,U|Z)\,\,=0=\,\diff h(Z,U|Y)\quad \Rightarrow\quad \ingl h(X,Y)\geq 0$,
\end{description}
which can be obtained from 
[\,2cI\,] and [\,4cI\,] by the exchange $[X,Z]\leftrightarrow [Y,U]$.
Each of 19 considered CI implications is derived from one of these 7 (versions of) conditional Ingleton inequalities
using one of 5 forms (M.1), (M.2), \ldots , (M.5) of the Ingleton expression from Lemma~\ref{lem.Inglet-forms}.

The derivations are in terms of the entropy function $h$ of a discrete
random vector over $XYZU$ satisfying the premises of the respective CI implication.
To describe them briefly we use special schematic records. Specifically,
the record
$$
\mbox{\rm (I:1)}\quad
\overbrace{\mbox{\footnotesize\rm [\,1cI\,]}} ~~~
\overbrace{\underline{(X,Y|\emptyset)}} \,+\, \overbrace{(X,Y|Z)}  \,+\, \underline{(Z,U|X)} \,+\,
\underline{(Z,U|Y)} ~~\stackrel{\mbox{\scriptsize (M.1)}}{\underline{~~}}~~
(Z,U|\emptyset)
$$
will be used to encode the following reasoning for (I:1). Its premise means
vanishing of four difference terms, encoded as the starting ones in the record:
$$
\diff h(X,Y|\emptyset)=0,\quad \diff h(X,Y|Z)=0,\quad \diff h(Z,U|X)=0,\quad  \diff h(Z,U|Y)=0,
$$
and the goal is to verify vanishing the remaining difference term in the record, which
is $\diff h(Z,U|\emptyset)=0$ in this case. One can use [\,1cI\,] inequality, which assumes
vanishing two terms $\diff h(X,Y|\emptyset)=0=\diff h(X,Y|Z)$, indicated by overbracing them in the record,
and says that $\ingl h(X,Y)\geq 0$. Then we use the (M.1) form of the Ingleton expression $\ingl h(X,Y)$,
which step is indicated by underlining in the record. Since the underlined terms vanish the Ingleton inequality
takes the form $-\diff h(Z,U|\emptyset)=\ingl h(X,Y)\geq 0$. Nevertheless, the remaining difference term
$\diff h(Z,U|\emptyset)$ is non-negative and must vanish.

The procedure to verify (I:2) is analogous, but uses [\,4cI\,] instead,
$$
\mbox{\rm (I:2)}\quad
\overbrace{\mbox{\footnotesize\rm [\,4cI\,]}} ~~~
\underline{(X,Y|\emptyset)} \,+\, \overbrace{(X,Z|U)}  \,+\, \overbrace{\underline{(Z,U|X)}} \,+\,
\underline{(Z,U|Y)} ~~\stackrel{\mbox{\scriptsize (M.1)}}{\underline{~~}}~~
(Z,U|\emptyset)
$$
with little extension: using the formula $\diff (Z,U|\emptyset)+\diff (X,Z|U)=\diff h(Z,XU|\emptyset)$
yields a stronger conclusion $Z\ci XU\,|\,\emptyset$.
\smallskip

The next bunch of CI implications is based on (M.2) form:
\begin{eqnarray*}
\mbox{\rm (I:3)}\quad
\overbrace{\mbox{\footnotesize\rm [\,cI2\,]}} ~~~
\underline{(X,Y|\emptyset)} \,+\, \overbrace{(X,Y|U)}  \,+\, \overbrace{\underline{(X,Z|U)}} \,+\,
\underline{(Z,U|Y)} ~~\stackrel{\mbox{\scriptsize (M.2)}}{\underline{~~}}~~
(X,Z|\emptyset) \\
\mbox{\rm (I:4)}\quad
\overbrace{\mbox{\footnotesize\rm [\,3cI\,]}} ~~~
\underline{(X,Y|\emptyset)} \,+\, \overbrace{\underline{(X,Z|U)}}  \,+\, \overbrace{(X,U|Z)} \,+\,
\underline{(Z,U|Y)} ~~\stackrel{\mbox{\scriptsize (M.2)}}{\underline{~~}}~~
(X,Z|\emptyset) \\
\mbox{\rm (I:5)}\quad
\overbrace{\mbox{\footnotesize\rm [\,cI4\,]}} ~~~
\underline{(X,Y|\emptyset)} \,+\, \underline{(X,Z|U)}  \,+\, \overbrace{(Y,U|Z)} \,+\,
 \underline{\overbrace{(Z,U|Y)}} ~~\stackrel{\mbox{\scriptsize (M.2)}}{\underline{~~}}~~
(X,Z|\emptyset) \\
\mbox{\rm (I:6)}\quad
\overbrace{\mbox{\footnotesize\rm [\,5cI\,]}} ~~~
\underline{(X,Y|\emptyset)} \,+\, \underline{\overbrace{(X,Z|U)}}  \,+\, \overbrace{(Y,Z|U)} \,+\,
\underline{(Z,U|Y)} ~~\stackrel{\mbox{\scriptsize (M.2)}}{\underline{~~}}~~
(X,Z|\emptyset)
\end{eqnarray*}
with an extension for (I:4): formula $\diff (X,Z|\emptyset)+\diff (X,U|Z)=\diff (X,ZU|\emptyset)$
yields a stronger conclusion $X\ci ZU\,|\,\emptyset$.
\smallskip

Another groups of CI implications is based on (M.3) form:
\begin{eqnarray*}
\mbox{\rm (I:7)}\quad
\overbrace{\mbox{\footnotesize\rm [\,1cI\,]}} ~~~
\overbrace{(X,Y|\emptyset)} \,+\, \underline{\overbrace{(X,Y|Z)}}  \,+\, \underline{(X,Z|U)} \,+\,
\underline{(Z,U|Y)} ~~\stackrel{\mbox{\scriptsize (M.3)}}{\underline{~~}}~~
(X,Z|Y) \\
\mbox{\rm (I:8)}\quad
\overbrace{\mbox{\footnotesize\rm [\,2cI\,]}} ~~~
\underline{\overbrace{(X,Y|Z)}} \,+\, \underline{(X,Z|U)}  \,+\, \overbrace{(Y,U|Z)} \,+\,
\underline{(Z,U|Y)} ~~\stackrel{\mbox{\scriptsize (M.3)}}{\underline{~~}}~~
(X,Z|Y) \\
\mbox{\rm (I:9)}\quad
\overbrace{\mbox{\footnotesize\rm [\,cI2\,]}} ~~~
\underline{(X,Y|Z)} \,+\, \overbrace{(X,Y|U)} \,+\, \overbrace{\underline{(X,Z|U)}}  \,+\,
 \underline{(Z,U|Y)} ~~\stackrel{\mbox{\scriptsize (M.3)}}{\underline{~~}}~~
(X,Z|Y) \\
\mbox{\rm (I:10)}\quad
\overbrace{\mbox{\footnotesize\rm [\,3cI\,]}} ~~~
\underline{(X,Y|Z)} \,+\, \underline{\overbrace{(X,Z|U)}}  \,+\, \overbrace{(X,U|Z)} \,+\,
\underline{(Z,U|Y)} ~~\stackrel{\mbox{\scriptsize (M.3)}}{\underline{~~}}~~
(X,Z|Y) \\
\mbox{\rm (I:11)}\quad
\overbrace{\mbox{\footnotesize\rm [\,4cI\,]}} ~~~
\underline{(X,Y|Z)} \,+\, \underline{\overbrace{(X,Z|U)}}  \,+\, \overbrace{(Z,U|X)} \,+\,
 \underline{(Z,U|Y)} ~~\stackrel{\mbox{\scriptsize (M.3)}}{\underline{~~}}~~
(X,Z|Y) \\
\mbox{\rm (I:12)}\quad
\overbrace{\mbox{\footnotesize\rm [\,5cI\,]}} ~~~
\underline{(X,Y|Z)} \,+\, \underline{\overbrace{(X,Z|U)}}  \,+\, \overbrace{(Y,Z|U)} \,+\,
\underline{(Z,U|Y)} ~~\stackrel{\mbox{\scriptsize (M.3)}}{\underline{~~}}~~
(X,Z|Y)
\end{eqnarray*}
with an extension for (I:7): $\diff (X,Z|Y)+\diff (X,Y|\emptyset)=\diff (X,YZ|\emptyset)$
yields a stronger conclusion $X\ci YZ\,|\,\emptyset$.
\smallskip

Two of CI implications are based on (M.4) form:
\begin{eqnarray*}
\mbox{\rm (I:13)}~~
\overbrace{\mbox{\footnotesize\rm [\,1cI\,]}} ~~
\overbrace{(X,Y|\emptyset)} \,+\, \underline{\overbrace{(X,Y|Z)}}  \,+\, \underline{(X,Y|U)} \,+\,
\underline{(Z,U|XY)} \stackrel{\mbox{\scriptsize (M.4)}}{\underline{~~}}
(X,Y|ZU) && ~~\\
\mbox{\rm (I:14)}~~
\overbrace{\mbox{\footnotesize\rm [\,cI2\,]}} ~~
\underline{(X,Y|Z)} \,+\, \underline{\overbrace{(X,Y|U)}}  \,+\, \overbrace{(X,Z|U)} \,+\,
\underline{(Z,U|XY)} \stackrel{\mbox{\scriptsize (M.4)}}{\underline{~~}}
(X,Y|ZU) && ~~
\end{eqnarray*}
with an extension for (I:14): $\diff (X,Y|ZU)+\diff (X,Z|U)=\diff (X,YZ|U)$
yields a stronger conclusion $X\ci YZ\,|\,U$.
\smallskip

The last bunch of CI implications is based on (M.5) form:
\begin{eqnarray*}
\,\mbox{\rm (I:15)}~~
\overbrace{\mbox{\footnotesize\rm [\,1cI\,]}} ~~
\overbrace{(X,Y|\emptyset)} \,+\, \underline{\overbrace{(X,Y|Z)}}  \,+\, \underline{(X,Z|U)} \,+\,
\underline{(Z,U|XY)} \stackrel{\mbox{\scriptsize (M.5)}}{\underline{~~}}
(X,Z|YU)&& ~~\\
\!\mbox{\rm (I:16)}~~\,
\overbrace{\mbox{\footnotesize\rm [\,2cI\,]}} ~~
\underline{\overbrace{(X,Y|Z)}} \,+\, \underline{(X,Z|U)}  \,+\, \overbrace{(Y,U|Z)} \,+\,
\underline{(Z,U|XY)} \stackrel{\mbox{\scriptsize (M.5)}}{\underline{~~}}
(X,Z|YU) && \\
\mbox{\rm (I:17)}~~
\overbrace{\mbox{\footnotesize\rm [\,3cI\,]}} ~~
\underline{(X,Y|Z)} \,+\, \underline{\overbrace{(X,Z|U)}} \,+\, \overbrace{(X,U|Z)}  \,+\,
 \underline{(Z,U|XY)} \stackrel{\mbox{\scriptsize (M.5)}}{\underline{~~}}
(X,Z|YU) &&\\
\mbox{\rm (I:18)}~~
\overbrace{\mbox{\footnotesize\rm [\,4cI\,]}} ~~
\underline{(X,Y|Z)} \,+\, \underline{\overbrace{(X,Z|U)}}  \,+\, \overbrace{(Z,U|X)} \,+\,
\underline{(Z,U|XY)} \stackrel{\mbox{\scriptsize (M.5)}}{\underline{~~}}
(X,Z|YU) &&\\
\mbox{\rm (I:19)}~~
\overbrace{\mbox{\footnotesize\rm [\,5cI\,]}} ~~
\underline{(X,Y|Z)} \,+\, \underline{\overbrace{(X,Z|U)}}  \,+\, \overbrace{(Y,Z|U)} \,+\,
 \underline{(Z,U|XY)} \stackrel{\mbox{\scriptsize (M.5)}}{\underline{~~}}
(X,Z|YU) &&
\end{eqnarray*}
with an extension for (I:19): $\diff (X,Z|YU)+\diff (Y,Z|U)=\diff (Z,XY|U)$
yields a stronger conclusion $Z\ci XY\,|\,U$.
\end{proof}

In the proof above we have derived each CI implication by combining two CI assumptions that imply Ingleton inequality with a suitable form of the Ingleton expression.
Note in this context that such a combination need not be unique, some of those
CI implications may have two alternative derivations which use different combinations
of these two things.

The reader willing to commemorate former occurrence of the properties from Corollary~\ref{cor.CI-implic}
can find detailed information in the Appendix, Section~\ref{sec.former-CI}.

\subsection{Sub-maximal CI structures}
The largest CI structure over $N$ is the one induced by a random vector
over $N$ whose components are stochastically independent random variables. In this section we discuss
{\em sub-maximal CI structures\/} induced by 4 discrete random variables (see Section~\ref{ssec.lattice} for definitions).
These appear to be closely related to the cone (of rank functions) of
polymatroids satisfying all six versions of Ingleton inequality. More specifically,
it was observed in \cite[Lemma~4.1]{MatStu95CPC} that, in case $|N|=4$, this pointed cone has 35 extreme rays
falling into 10 permutational types. In our context of standard CI structures, however, one can limit the attention
to the sub-cone of that cone specified by additional requirements $\diff (i,i|N\setminus i)=0$ for $i\in N$; this leads to excluding 4 extreme rays of 1 permutational type.

The polymatroids generating the extreme rays of the sub-cone mentioned above
are linearly representable, and, therefore, by utilizing the results from \cite{Pad12},
can be shown to be probabilistically representable.
The induced CI structures then appear to be sub-maximal CI structures induced by 4 discrete
random variables. To give the reader a complete picture we list representatives of 9
permutational types of these CI structures in the Appendix, Section~\ref{sec.sub-max-constr}.

\subsection{Overview: two complementary characterizations}
In this section we summarize our findings on (standard) CI structures induced by 4 discrete random variables.
Since these structures are semi-graphoids it follows from the facts in Section~\ref{ssec.semi-graph} that
they can be identified with subsets of a 24-element set of (pairs of mutually equivalent) elementary triplets
$(i,j|K)$ over $N$, $|N|=4$ (see Lemma~\ref{lem.local-semi}). Note in this context that the result of computations
made in connection with \cite{Stu94IJGS} was that the number of (standard) semi-graphoids over $N$, $|N|=4$,
is 26\,424. An even tighter upper approximation is the set of ``structural" semi-graphoids over $N$ (see Remark~\ref{rem.structural}); their number is 22\,108 \cite{Stu94IJGS}. The number of CI structures induced by 4 discrete random variables is, however, only 18\,478, which is the result of computations made in connection with \cite{Sim07}.
\medskip

The class of CI structures is a lattice because the intersection of two CI structures is a CI structure
(see Section~\ref{ssec.lattice}). This fact leads to two mutually complementary ways to characterize the CI structures, namely
\begin{itemize}
\item in terms of irreducible CI structures, and
\item in terms of CI implications.
\end{itemize}
The first option is straightforward:  to characterize CI structures over $N$, \mbox{$|N|=4$,} it is enough to provide the list
(of permutational types) of meet-irreducible CI structures over $N$. These appear to be
\begin{itemize}
\item the full CI structure (induced by independent random variables),
%
\item the sub-maximal CI structures (= the co-atoms of the lattice),
which are described in the Appendix, Section~\ref{sec.sub-max-constr},
\item remaining meet-irreducible CI structures, which are just those presented
in Examples~\ref{exa.M-1}-\ref{exa.M-4} from Section~\ref{ssec.counter-Ingleton}.
\end{itemize}
Note that the irreducible CI structures of the third kind correspond to 
counter-examples to probabilistic validity of Ingleton inequality. Thus, altogether one has 92 irreducible discrete probabilistic CI structures over $N$, $|N|=4$,
and they break into 14 permutational types.
\smallskip

The second option is to describe minimal sets ${\cal X}$ of (elementary) triplets over $N$ that
{\em are not\/} CI structures.
Every such a set has uniquely determined CI closure: the fact that the intersection of CI structures
is a CI structure implies the existence of the least CI structure ${\cal Y}$ (over $N$) containing ${\cal X}$.
One can interpret this as the claim that ${\cal X}$ probabilistically implies
${\cal Y}\setminus {\cal X}$. Therefore, the above mentioned minimal sets ${\cal X}$
correspond to CI implications/equivalences. One can distinguish three kinds of these
CI properties in case $|N|=4$:
\begin{itemize}
\item the semi-graphoid properties (S:0)-(S:2) discussed in Section~\ref{ssec.semi-graph},
\item the CI equivalences (E:1)-(E:5) from Corollary~\ref{cor.CI-equiv},
\item the CI implications (I:1)-(I:19) from Corollary~\ref{cor.CI-implic}.
\end{itemize}
Thus, altogether, one can recognize 27 abstract CI properties characterizing
CI structures induced by 4 discrete random variables. The fact that both
ways of description determine the same class of 18\,478 (standard) abstract CI
structures over $N$, $|N|=4$, was verified by means of a computer
\cite{Sim07}.

\section{Conclusions}\label{sec.conclusions}
The principal message of the paper is that CI inference problem (for discrete
probability distributions) can be solved in an elegant way by applying
special conditional information inequalities. This was testified in case in case of four random variables
and the question is whether the idea is applicable in general, for a higher number of random variables.

A new conditional Ingleton inequality was revealed in this paper and this leads to
a few natural open tasks formulated in Section~\ref{ssec.open-Ingleton}. Note
in this context that conditional information inequalities  have applications outside information theory.
For example, (more general) conditional information inequalities can be used to obtain bounds for
information ratios in the area of secrete sharing (schemes) \cite[\S\,3.2.6]{AKNS20IEEE}, have some group-theoretical
counter-parts \cite[\S\,3.2.4]{AKNS20IEEE} and also counter-parts in context of Kolgomorov complexity
\cite[\S\,7]{KacRom13IEEE}. One can even find combinatorial applications for conditional Ingleton
inequalities \cite{KRV18IEEE}.

This paper has not dealt with augmented CI structures involving functional dependence statements. For this reason, the conditional Ingleton inequalities which have functional dependence premises allowed, as for example \mbox{(${\cal I}3)$} in \cite{KacRom13IEEE}, have been omitted.
These inequalities can, however, possibly be used to derive (further) CI implications for the augmented
probabilistic CI structures. Thus, one of future potential research directions is to
(try to) characterize augmented CI structures induced by 4 discrete random variables by means of
conditional information inequalities allowing functional dependence premises.

The interest in (algorithmic aspects of) constraints on entropy functions has recently
been amplified in the context of database theory \cite{AKNS20IEEE}. An interesting
result from that paper, related to our topic of study, says that the CI inference problem (with a bounded number of variables) belongs to one of primitive class of decidable problems \cite[Theorem\,8]{AKNS20IEEE}.
The results presented in this paper support that claim from \cite{AKNS20IEEE} and raise the hope in characterizing CI inference for a higher number of discrete random variables.


\subsubsection*{Acknowledgements}
The research of the author is supported from GA\v{C}R project 19-04579S.
This particular research theme follows up the cooperation with my deceased colleague Fero Mat\'{u}\v{s} and I am
indebted to him for the fruitful discussion on this topic we had over the former years. My thanks also go to Tobias Boege, who reminded me that every linearly representable matroid is representable over a finite field.
Finally, I am indebted to L\'{a}szl\'{o} Csirmaz, Tarik Kaced, and anonymous reviewers, for reading the manuscript and giving me their valuable comments.




\appendix

\section{Appendix: proofs of inequalities}\label{sec.appendix-proofs}
We prove all five conditional Ingleton inequalities from Section~\ref{sec.condi-Ingleton} here.

\subsection{The first inequality} 
The first conditional Ingleton inequality appeared as \cite[Theorem~3]{ZhaYeu97ITI}
but it was not immediately clear from its original formulation that it is, in fact, the Ingleton inequality.
For sake of completeness we present a modified version of the original
proof from \cite[Section\,II]{ZhaYeu97ITI}; it seems to be the simplest possible proof
of this inequality from a methodological point of view.

\begin{proposition}\em
The first conditional Ingleton inequality is valid:
\begin{description}
\item[\mbox{[\,1cI\,]}] ~~ $\diff(X,Y|\emptyset)=0=\diff(X,Y|Z)\,\,\quad \Rightarrow\quad \ingl(X,Y)\geq 0$.
\end{description}
\end{proposition}

\begin{proof}
Let $P$ be a probability distribution over $XYZU$ that satisfies both
$X\ci Y\,|\,\emptyset\,\,[P]$ and $X\ci Y\,|\,Z\,\,[P]$, which is equivalent to the
assumption that the entropy function $h$ of $P$ satisfies $\diff h(X,Y|\emptyset)=0=\diff h(X,Y|Z)$.
We are going to show $\ingl h(X,Y)\geq 0$.

Let $Q$ be the conditional product of marginals $P_{XZU}$ and $P_{YZU}$ (see Section~\ref{ssec.cond-prod})
and $\overline{h}$ the
entropy function of $Q$. Thus, $P$ and $Q$ have the same marginals for $XZU$ and $YZU$ while one has,
moreover, $X\ci Y\,|\,ZU\,\,[Q]$.
The next step is to define a probability distribution $R$ over $XYZU$ which
shares the (joint) sample space with $P$ and $Q$. Specifically, its density $r$ is given by
$$
r(x,y,z,u) ~:=~
\left\{
\begin{array}{cl}
\,\frac{p(x,z)\cdot p(x,u)\cdot p(y,z)\cdot p(y,u)}{p(x)\cdot p(y)\cdot p(z)\cdot p(u)}
& ~~\mbox{if $p(x)\cdot p(y)\cdot p(z)\cdot p(u)>0$,}\\[2ex]
\,0 & ~~\mbox{otherwise,}
\end{array}
\right.
$$
where $x,y,z,u$ belong to the respective individual sample spaces and $p(x,z)$ denotes
the value of the marginal density of $P$ for $XZ$ in the configuration $[x,z]$.
Note that, in the definition of $r$, one can replace the used marginal densities $p(*)$ of $P$ by
the respective marginal densities $\overline{p}(*)$ of $Q$ because they coincide.
Our CI assumptions imply that $r$ is a probability density:
\begin{eqnarray*}
\lefteqn{\hspace*{-5mm}\sum_{x,y,z,u} r(x,y,z,u) = \sum_{x,y,z,u:\,  p(x),\ldots,p(u)>0} \frac{p(x,z)\cdot p(x,u)\cdot p(y,z)\cdot p(y,u)}{p(x)\cdot p(y)\cdot p(z)\cdot p(u)} }\\
&\stackrel{X\ici Y|Z\,[P]}{=}& \sum_{x,y,z,u:\,  p(x),\ldots,p(u)>0}\, \frac{p(x,y,z)\cdot p(x,u)\cdot p(y,u)}{p(x)\cdot p(y)\cdot p(u)}\\
 &=&
\sum_{x,y,u:\,p(x),p(y),p(u)>0}\, \frac{p(x,u)\cdot p(y,u)}{p(x)\cdot p(y)\cdot p(u)}\cdot
\underbrace{\sum_{z:\,p(z)>0}\, p(x,y,z)}_{p(x,y)}\\
 &=&
\sum_{x,y,u:\,p(x),p(y),p(u)>0}\, \frac{p(x,u)\cdot p(y,u)\cdot p(x,y)}{p(x)\cdot p(y)\cdot p(u)}\\
&\stackrel{X\ici Y|\emptyset\,[P]}{=}& \sum_{x,y,u:\,p(x),p(y),p(u)>0}\, \frac{p(x,u)\cdot p(y,u)}{p(u)}\\
&=& \sum_{x,u:\,p(x),p(u)>0}\, \frac{p(x,u)}{p(u)}\cdot \underbrace{\sum_{y:\,p(y)>0} p(y,u)}_{p(u)}\\
&=& \sum_{x,u:\,p(x),p(u)>0}\, p(x,u) ~=~ 1\,.
\end{eqnarray*}
The definition of $R$ also implies that $Q$ is absolutely continuous with respect to $R$ because
$r(x,y,z,u)=0$ implies $p(x,z)\cdot p(x,u)\cdot p(y,z)\cdot p(y,u)=0$ and, hence,
$\overline{p}(x,y,z,u)=0$, where $\overline{p}$ is the density of $Q$. It remains to verify that
$\ingl h(X,Y)=D(Q\,\|\, R)$ to which goal we use $X\ci Y\,|\,\emptyset\,\,[P]$ and  $X\ci Y\,|\,ZU\,\,[Q]$:
\begin{eqnarray*}
\lefteqn{\hspace*{-1mm}\ingl h(X,Y) \stackrel{\mbox{\scriptsize (M.1)}}{=} \diff h (Z,U|X) +  \diff h (Z,U|Y)
+\underbrace{\diff h (X,Y|\emptyset)}_{=0} - \diff h (Z,U|\emptyset)}\\
&=& \diff \overline{h} (Z,U|X) +  \diff \overline{h} (Z,U|Y) - \diff \overline{h} (Z,U|\emptyset)\\
&=& -\,\overline{h}(XZU)-\overline{h}(YZU)+\overline{h}(ZU)\\
&& \,\,+\,\overline{h}(XZ)+\overline{h}(XU)+\overline{h}(YZ)+\overline{h}(YU)-\overline{h}(X)-\overline{h}(Y)-\overline{h}(Z)-\overline{h}(U)\\
&=& \sum_{x,y,z,u\,:\, \overline{p}(x,y,z,u)>0} \overline{p}(x,y,z,u)\cdot \ln\,
\frac{\frac{\overline{p}(x,z,u)\cdot \overline{p}(y,z,u)}{\overline{p}(z,u)}}{\frac{\overline{p}(x,z)\cdot \overline{p}(x,u)\cdot \overline{p}(y,z)\cdot \overline{p}(y,u)}{\overline{p}(x)\cdot \overline{p}(y)\cdot \overline{p}(z)\cdot \overline{p}(u)}}\\
&=& \sum_{x,y,z,u\,:\, \overline{p}(x,y,z,u)>0} \overline{p}(x,y,z,u)\cdot \ln\,
\frac{\overline{p}(x,y,z,u)}{r(x,y,z,u)} ~=~ D(Q\,\|\, R)\geq 0\,.
\end{eqnarray*}
Thus, the non-negativity of the Kullback-Leibler divergence implies the claim.
\end{proof}

\subsection{The second inequality} 
The second conditional Ingleton inequality appeared in \cite[\S\,2]{Mat99CPC}.
Specifically, the proof of Proposition~2.1 in \cite{Mat99CPC} contained both a hint to verify the first inequality
from \cite{ZhaYeu97ITI} and the arguments for the second inequality. We believe that our proof of the second inequality
is much simpler than the original one.

\begin{proposition}\em
The second conditional Ingleton inequality is valid:
\begin{description}
\item[\mbox{[\,2cI\,]}] ~~ $\diff(X,Y|Z)=0=\diff(Y,U|Z)\quad \Rightarrow\quad \ingl(X,Y)\geq 0$.
\end{description}
\end{proposition}

\begin{proof}
Let $P$ be a probability distribution over $XYZU$ that satisfies both
$X\ci Y\,|\,Z\,\,[P]$ and $Y\ci U\,|\,Z\,\,[P]$, which is equivalent to the
assumption that the entropy function $h$ of $P$ satisfies $\diff h(X,Y|Z)=0=\diff h(Y,U|Z)$.
We are going to show $\ingl h(X,Y)\geq 0$.

Let $Q$ be the conditional product of marginals $P_{YZ}$ and $P_{XZU}$ and $\overline{h}$ the
entropy function of $Q$. Thus, $P$ and $Q$ have the same marginals for $YZ$ and $XZU$ while one,
moreover, has $Y\ci XU\,|\,Z\,\,[Q]$. Hence, $Y\ci U\,|\,Z\,\,[Q]$ and the facts
$P_{YZ}=Q_{YZ}$, $P_{ZU}=Q_{ZU}$ and $Y\ci U\,|\,Z\,\,[P]$ imply $P_{YZU}=Q_{YZU}$.
One can analogously derive $P_{XYZ}=Q_{XYZ}$ (exchange $U$ and $X$).
In particular, $P$ and $Q$ have the same marginals for $XZU$, $YZU$, and $XY$ and it follows
from the definition \eqref{eq.Inglet} of the Ingleton expression that $\ingl h(X,Y)=\ingl \overline{h}(X,Y)$.
Therefore, it is enough to show $\ingl \overline{h}(X,Y)\geq 0$.

Nonetheless, $Y\ci XU\,|\,Z\,\,[Q]$ implies $Y\ci X\,|\,ZU\,\,[Q]$, which is equivalent to
$\diff \overline{h} (X,Y|ZU)=0$.
Analogously, $\diff \overline{h} (X,Y|Z)=0$ can be observed, although this particular observation
is not necessary to draw our conclusion.
Thus, one of the forms of the Ingleton expression from Lemma~\ref{lem.Inglet-forms}
\begin{eqnarray*}
\lefteqn{\hspace*{-21mm}\ingl \overline{h}(X,Y) \stackrel{\mbox{\scriptsize (M.4)}}{=}
\underbrace{\diff \overline{h} (X,Y|Z)}_{=0} + \diff \overline{h} (X,Y|U)
+ \diff \overline{h} (Z,U|XY) - \underbrace{\diff \overline{h} (X,Y|ZU)}_{=0}} \\
&=& \diff \overline{h} (X,Y|U) + \diff \overline{h} (Z,U|XY) \hspace*{2cm}~~
\end{eqnarray*}
together with Shannon inequalities for $\overline{h}$ imply that $\ingl \overline{h}(X,Y)\geq 0$.
\end{proof}

\subsection{The third inequality}\label{ssec.app.3rd} 
A number of CI implications has been derived in \cite[\S\,2]{Mat95CPC} by a special technique
of transforming the implication problem into a wider framework of $\sigma$-algebras and applying
a special CI property for $\sigma$-algebras based on a construction of a certain
``\,intersection" $\sigma$-algebra for two given $\sigma$-algebras. The idea behind this method is that random
variables on a probability space $(\Omega,{\cal A},\boP)$ can be identified with their induced
sub-$\sigma$-algebras of ${\cal A}$; yet another interpretation is possible
in case of a finite $\Omega$ when the $\sigma$-algebras correspond to partitions of $\Omega$.

In our case of finitely many discrete random variables,  the construction of the ``\,intersection"
$\sigma$-algebra corresponds to a construction of a certain random variable on basis of two given
random variables. The above special CI property for $\sigma$-algebras can equivalently be expressed as the property of the constructed random variable, which is formally done in the next Lemma~\ref{lem.intersec}.
Note that this auxiliary result has already been formulated both in \cite[Lemma\,1]{KacRom13IEEE} and in \cite[Lemma\,4]{MMRV02}, where it was named a ``double Markov property".
Nevertheless, the result itself was known much earlier: it was already formulated in the form of an exercise
under title ``double Markovity" in \cite[\S\,3.4, excercise~25]{CK86}.

\begin{lemma}\em \label{lem.intersec}
Every discrete random vector $\bxi$ over $ABC$ satisfying $\xi_{A}\ci \xi_{B}\,|\,\xi_{C}$
and $\xi_{A}\ci \xi_{C}\,|\,\xi_{B}$ can be extended by a discrete random variable $\xi_{W}$
to a random vector over $ABCW$ in which $\xi_{W}$ is functionally dependent both on $\xi_{B}$ and on $\xi_{C}$
and, moreover, one has $\xi_{A}\ci \xi_{BC}\,|\,\xi_{W}$.
\end{lemma}

\begin{proof}
Assume without loss of generality $A,B,C\neq\emptyset$ for otherwise one can take
any constant random variable in place of $\xi_{W}$.
Let $P$ be the distribution of $\bxi$ on a joint finite sample space ${\sf A}\times {\sf B}\times {\sf C}$
and $p$ its density with strictly positive one-dimensional marginal densities on ${\sf A}$, ${\sf B}$, and ${\sf C}$.
Consider the support set ${\sf L}:=\{\, (b,c)\in {\sf B}\times {\sf C}\,:\ p(b,c)>0\,\}$
of the marginal distribution of $P$ for $BC$ equipped with a binary relation $\sim$ defined by
$$
(b,c)\sim (b^{\prime},c^{\prime}) ~:=~ [\, b=b^{\prime} ~~\mbox{or}~~ c=c^{\prime}\,]
$$
and introduce an equivalence $\approx$ on ${\sf L}$ as the transitive closure of\/ $\sim$\,.
Then the quotient set ${\sf W}$ of ${\sf L}$ by\/ $\approx$ (= the collection of equivalence classes for $\approx$)
will be the individual sample space for\/ $\xi_{W}$. Define the density $\tilde{p}$ of the
extended random vector over $ABCW$ as follows: given $(a,b,c,w)\in {\sf A}\times {\sf B}\times {\sf C}\times {\sf W}$,
$$
\tilde{p}(a,b,c,w) ~:=~
\left\{
\begin{array}{cl}
\,p(a,b,c)
& ~~\mbox{if $p(a,b,c)>0$ and $w$ contains $(b,c)$,}\\[1ex]
\,0 & ~~\mbox{otherwise.}
\end{array}
\right.
$$
Given $b\in {\sf B}$, the definition of\/ $\approx$ implies that
all pairs $(b,c)\in {\sf L}$ belong to the same equivalence class $w$ of\/ $\approx$.
Thus, $\xi_{W}$ functionally depends on\/ $\xi_{B}$ and an analogous argument implies
that $\xi_{W}$ functionally depends on\/ $\xi_{C}$.

Realize that $P$ can be viewed as a distribution on ${\sf A}\times {\sf L}$ and
one can consider the conditional density $p_{A|BC}(a\,|\,b,c):= \frac{p(a,b,c)}{p(b,c)}$
defined for $a\in{\sf A}$ and $(b,c)\in {\sf L}$. The assumption
$\xi_{A}\ci \xi_{B}\,|\,\xi_{C}$ implies
$p_{A|BC}(a\,|\, b,c)=\frac{p(a,c)}{p(c)}=p_{A|BC}(a\,|\, b^{\prime},c)$
whenever $a\in {\sf A}$ and $(b,c),(b^{\prime},c)\in {\sf L}$.
Analogously, the assumption $\xi_{A}\ci \xi_{C}\,|\,\xi_{B}$ gives
$p_{A|BC}(a\,|\, b,c)=p_{A|BC}(a\,|\, b,c^{\prime})$ whenever $a\in {\sf A}$
and $(b,c),(b,c^{\prime})\in {\sf L}$. Hence, $p_{A|BC}(\,*\,|\, b,c)=p_{A|BC}(\,*\,|\, b^{\prime},c^{\prime})$
whenever $(b,c)\sim (b^{\prime},c^{\prime})$ and, by transitivity argument, $p_{A|BC}$ is constant on equivalence
classes of\/ $\approx$. Thus, having fixed an equivalence class $w\in {\sf W}$, the conditional density
$p_{A|BC}(\,*\,|\, b,c)$ does not depend on the choice of $(b,c)\in w\subseteq {\sf L}$, which implies
$\xi_{A}\ci \xi_{BC}\,|\,\xi_{W}$.
\end{proof}

The random variable $\xi_{W}$ from the above proof is, in fact, defined by
some equivalence\/ $\approx$ on ${\sf L}\subseteq {\sf B}\times{\sf C}$. The particular definition
of\/ $\approx$ from our proof corresponds to the construction of the ``\,intersection" $\sigma$-algebra
for $\sigma$-algebras induced by $\xi_{B}$ and $\xi_{C}$; note that it solely depends on the marginal of $P$ on $BC$.
Thus, our construction is universal relative to $\xi_{A}$ in sense that one gets $\xi_{A}\ci \xi_{BC}\,|\,\xi_{W}$
for {\em arbitrary\/} $\xi_{A}$ satisfying $\xi_{A}\ci \xi_{B}\,|\,\xi_{C}$ and $\xi_{A}\ci \xi_{C}\,|\,\xi_{B}$.

Note in this context that the proof of an analogous result in \cite[Appendix\,IX]{KacRom13IEEE} was based on a slightly different, perhaps formally shorter, construction of an equivalence on ${\sf L}$,
which construction was, however, depending on the considered random variable $\xi_{A}$. Thus, its specific form was
determined by the marginal of $P$ on $ABC$. In particular, the equivalence on ${\sf L}$ from
\cite[Appendix\,IX]{KacRom13IEEE} can be shown to be a coarsening of\/ $\approx$ from our proof of Lemma~\ref{lem.intersec}.
\smallskip

We now apply Lemma~\ref{lem.intersec} to prove the third conditional Ingleton inequality
which appeared in \cite{KacRom13IEEE} as \mbox{(${\cal I}4^{\prime})$}.
In fact, two proofs of this inequality were given there. The basic proof of it was more
complicated because the goal was to prove that it is valid for any function which is a limit of
(discrete) entropy functions. The second ``direct" proof from \cite[\S~III.C]{KacRom13IEEE} was
simpler because of easier task to prove that the inequality holds for entropy functions only.
That simpler proof was based on an analogous idea as our proof below but it was different in technical details.
We believe that our reasoning is more transparent.

\begin{proposition}\em
The third conditional Ingleton inequality is valid:
\begin{description}
\item[\mbox{[\,3cI\,]}]~~ $\diff(X,Z|U)=0=\diff(X,U|Z)\quad \Rightarrow\quad \ingl(X,Y)\geq 0$.
\end{description}
\end{proposition}

\begin{proof}
Let $P$ be a probability distribution over $XYZU$ that satisfies both
$X\ci Z\,|\,U\,\,[P]$ and $X\ci U\,|\,Z\,\,[P]$, which is equivalent to the
assumption that the entropy function $h$ of $P$ satisfies $\diff h(X,Z|U)=0=\diff h(X,U|Z)$.
We are going to show $\ingl h(X,Y)\geq 0$.

We apply Lemma~\ref{lem.intersec} to a random vector $\bxi$ over $XZU$ having $P_{XZU}$ as its distribution
with $A:=X$, $B:=Z$ and $C:=U$. It says there is an extended discrete random vector $\tilde{\bxi}$ over $XZUW$ with
a distribution $\tilde{P}$ (having $P_{XZU}$ as a marginal) such that the variable $\tilde{\xi}_{W}$ functionally depends both
on $\tilde{\xi}_{Z}$ and on $\tilde{\xi}_{U}$ and, moreover, $\tilde{\xi}_{X}\ci \tilde{\xi}_{ZU}\,|\,\tilde{\xi}_{W}$.
Let us introduce the conditional product $\hat{P}$ of distributions $\tilde{P}$ over $XZUW$ and $P$ over $XYZU$, that is,
define an extended random vector $\hat{\bxi}$ over $XYZUW$. By the construction, its entropy
function $\hat{h}$ is an extension of the entropy function $h$ of $P$, one has $X\ci ZU\,|\,W\,\,[\hat{P}]$
and $\hat{\xi}_{W}$ functionally depends both on $\hat{\xi}_{Z}$ and on $\hat{\xi}_{U}$.
\smallskip

To verify the conclusion we use the dependencies $\hat{\xi}_{W}\leftarrow\hat{\xi}_{Z}$ and $\hat{\xi}_{W}\leftarrow\hat{\xi}_{U}$,
which imply that $\hat{h}(T)=\hat{h}(TW)$ for any $T$ containing either $Z$ or $U$,
the fact that the entropy function is non-decreasing, which gives $\hat{h}(XY)\leq \hat{h}(XYW)$,
and $X\ci Z\,|\,W\,\,[\hat{P}]$, which gives $\diff \hat{h}(X,Z|W)=0$:
\begin{eqnarray*}
\lefteqn{\hspace*{-1mm}\ingl h(X,Y) \stackrel{\mbox{\scriptsize (M.2)}}{=} \diff h (Z,U|Y)
+ \underbrace{\diff h (X,Z|U)}_{=0}
 \,+\, \diff h (X,Y|\emptyset) - \diff h (X,Z|\emptyset)}\\
&=& \diff \hat{h} (Z,U|Y) + \diff  \hat{h} (X,Y|\emptyset) - \diff \hat{h} (X,Z|\emptyset)\\
&=& \hat{h}(YZ) + \hat{h}(YU) - \hat{h}(YZU) - \hat{h}(XY) - \hat{h}(Z) + \hat{h}(XZ) \\
&=& \hat{h}(YZW) + \hat{h}(YUW) - \hat{h}(YZUW) - \hat{h}(XY) - \hat{h}(ZW) + \hat{h}(XZW)\\
&\geq & \hat{h}(YZW) + \hat{h}(YUW) - \hat{h}(YZUW) - \hat{h}(XYW) - \hat{h}(ZW) + \hat{h}(XZW) \\
&=&  \diff \hat{h} (Z,U|YW) + \diff  \hat{h} (X,Y|W) - \underbrace{\diff \hat{h} (X,Z|W)}_{=0}\\
&=& \diff \hat{h} (Z,U|YW) + \diff  \hat{h} (X,Y|W)\geq 0\,.
\end{eqnarray*}
Thus, the Shannon's inequalities for $\hat{h}$ imply the desired conclusion.
\end{proof}

\subsection{The fourth inequality} 
Here we use an analogous method to verify the fourth conditional Ingleton inequality
which appeared as \mbox{(${\cal I}5^{\prime})$} in \cite{KacRom13IEEE}. Its original proof
was more intricate because the task was to prove its validity for functions which are limits of
entropy functions. Our proof below is based on different simpler arguments.

\begin{proposition}\em
The fourth conditional Ingleton inequality is valid:
\begin{description}
\item[\mbox{[\,4cI\,]}]~~ $\diff(X,Z|U)=0=\diff(Z,U|X) \quad\Rightarrow\quad \ingl(X,Y)\geq 0$.
\end{description}
\end{proposition}

\begin{proof}
Let $P$ be a probability distribution over $XYZU$ that satisfies both
$Z\ci X\,|\,U\,\,[P]$ and $Z\ci U\,|\,X\,\,[P]$, which is equivalent to the
assumption that the entropy function $h$ of $P$ satisfies $\diff h(X,Z|U)=0=\diff h(Z,U|X)$.
We are going to show $\ingl h(X,Y)\geq 0$.

We apply Lemma~\ref{lem.intersec} to a random vector $\bxi$ over $XZU$ having $P_{XZU}$ as its distribution
with $A:=Z$, $B:=X$ and $C:=U$. It says there is an extended discrete random vector $\tilde{\bxi}$ over $XZUW$ with
a distribution $\tilde{P}$ (having $P_{XZU}$ as a marginal) such that the variable $\tilde{\xi}_{W}$ functionally depends both on $\tilde{\xi}_{X}$ and on $\tilde{\xi}_{U}$ and, moreover,
$\tilde{\xi}_{Z}\ci \tilde{\xi}_{XU}\,|\,\tilde{\xi}_{W}$.
Take the conditional product $\hat{P}$ of distributions $\tilde{P}$ over $XZUW$ and $P$ over $XYZU$, that is,
define an extended random vector $\hat{\bxi}$ over $XYZUW$. Because of the construction, its entropy
function $\hat{h}$ is an extension of the entropy function $h$ of $P$, one has $Z\ci XU\,|\,W\,\,[\hat{P}]$
and $\hat{\xi}_{W}$ functionally depends both on $\hat{\xi}_{X}$ and on $\hat{\xi}_{U}$.
\smallskip

To verify the inequality we use the dependencies $\hat{\xi}_{W}\leftarrow\hat{\xi}_{X}$ and $\hat{\xi}_{W}\leftarrow\hat{\xi}_{U}$,
which imply that $\hat{h}(T)=\hat{h}(TW)$ for any $T$ containing either $X$ or $U$, and $Z\ci X\,|\,W\,\,[\hat{P}]$,
which gives $\diff \hat{h}(Z,X|W)=0$:
\begin{eqnarray*}
\lefteqn{\hspace*{-5mm}\ingl h(X,Y) \stackrel{\mbox{\scriptsize (M.2)}}{=} \diff h (Z,U|Y)
+ \underbrace{\diff h (X,Z|U)}_{=0}
+ \diff h (X,Y|\emptyset) - \diff h (X,Z|\emptyset)}\\
&=& \diff \hat{h} (Z,U|Y) +\diff \hat{h} (X,Y|\emptyset) - \diff \hat{h} (X,Z|\emptyset)\\
&=& \hat{h}(YZ) + \hat{h}(YU) - \hat{h}(YZU) - \hat{h}(XY) - \hat{h}(Z) + \hat{h}(XZ) \\
&=& \hat{h}(YZ) + \hat{h}(YUW) - \hat{h}(YZUW) - \hat{h}(XYW)- \hat{h}(Z) + \hat{h}(XZW)\\
&=& \hat{h}(YUW) - \hat{h}(YZUW) - \hat{h}(XYW) + \hat{h}(YZ) - \hat{h}(Z) + \hat{h}(XZW)\\
&=&  \diff \hat{h} (Z,U|YW) + \diff  \hat{h} (X,Y|W)  + \diff \hat{h} (Y,W|Z) -
\underbrace{\diff \hat{h} (Z,X|W)}_{=0}\\
&=& \diff \hat{h} (Z,U|YW) + \diff  \hat{h} (X,Y|W)  + \diff \hat{h} (Y,W|Z)\geq 0\,.
\end{eqnarray*}
\normalsize
Thus, the Shannon's inequalities for $\hat{h}$ imply the desired conclusion.
\end{proof}

\subsection{The fifth inequality}
Here we give the proof of the fifth conditional Ingleton inequality
which is methodologically similar to the proof of the second inequality.

\begin{proposition}\em
The fifth conditional Ingleton inequality is valid:
\begin{description}
\item[\mbox{[\,5cI\,]}]~~ $\diff(X,Z|U)=0=\diff(Y,Z|U)\quad \Rightarrow\quad \ingl(X,Y)\geq 0$.
\end{description}
\end{proposition}

\begin{proof}
Let $P$ be a probability distribution over $XYZU$ that satisfies both
$Z\ci X\,|\,U\,\,[P]$ and $Z\ci Y\,|\,U\,\,[P]$, which is equivalent to the
assumption that the entropy function $h$ of $P$ satisfies $\diff h(X,Z|U)=0=\diff h(Y,Z|U)$.
We are going to show $\ingl h(X,Y)\geq 0$.

Let $Q$ be the conditional product of marginals $P_{XYU}$ and $P_{ZU}$ and $\overline{h}$ the
entropy function of $Q$. Thus, $P$ and $Q$ have the same marginals for $XYU$ and $ZU$ while one,
moreover, has $Z\ci XY\,|\,U\,\,[Q]$. Hence, $Z\ci X\,|\,U\,\,[Q]$ and the facts
$P_{ZU}=Q_{ZU}$, $P_{XU}=Q_{XU}$ and $Z\ci X\,|\,U\,\,[P]$ imply $P_{XZU}=Q_{XZU}$.
Analogously, $P_{YZU}=Q_{YZU}$ (exchange $X$ and $Y$).
In particular, $P$ and $Q$ have the same marginals for $XZU$, $YZU$, and $XY$ and it follows
from the definition \eqref{eq.Inglet} of the Ingleton expression that $\ingl h(X,Y)=\ingl \overline{h}(X,Y)$.
Therefore, it is enough to show $\ingl \overline{h}(X,Y)\geq 0$.

Nonetheless, $Z\ci XY\,|\,U\,\,[Q]$ implies $Z\ci X\,|\,YU\,\,[Q]$, which is equivalent to $\diff \overline{h} (X,Z|YU)=0$.
Analogously, $\diff \overline{h} (X,Z|U)=0$ can be observed, although this particular observation is not necessary to draw our conclusion.
Thus, one of the forms of the Ingleton expression from Lemma~\ref{lem.Inglet-forms}
\begin{eqnarray*}
\lefteqn{\hspace*{-20.5mm}\ingl \overline{h}(X,Y) \stackrel{\mbox{\scriptsize (M.5)}}{=} \diff \overline{h} (X,Y|Z)
+ \underbrace{\diff \overline{h} (X,Z|U)}_{=0} + \diff \overline{h} (Z,U|XY)
- \underbrace{\diff \overline{h} (X,Z|YU)}_{=0}} \\
&=& \diff \overline{h} (X,Y|Z) + \diff \overline{h} (Z,U|XY) \hspace*{2cm}~~
\end{eqnarray*}
together with Shannon inequalities for $\overline{h}$ imply that $\ingl \overline{h}(X,Y)\geq 0$.
\end{proof}

\section{Appendix: sub-maximal CI structures}\label{sec.sub-max-constr}
This is to recall basic constructions of discrete
distributions over $N$, $|N|=4$.

\paragraph{\fbox{I.}}
Put $N=\{x,y,z,u\}$ and define the density $p$ of a binary random vector:
\begin{center}
\begin{tabular}{lclc}
& $x~y~z~u$ &&\\
$p:$ & $(\0 ,\0 ,\0 ,\0 )$ &$\mapsto$ & $1/2$ \\
& $(\1 ,\1 ,\0 ,\0 )$ &$\mapsto$ & $1/2$
\end{tabular}
\quad with zero values for other configurations.
\end{center}
It provides (strong) probabilistic representation of a polymatroid over $N$ with
the following integer-valued rank function:
$$
h(S) ~:=~ \min\,\{\,|S\cap\{x,y\}|\,,\,1\,\}
\quad \mbox{for any $S\subseteq N$.}
$$
Note it corresponds to the uniform matroid of rank 1 over $\{x,y\}$.
The induced CI structure 
involves 20 (pairs of) elementary CI statements:
\begin{eqnarray*}
& x\,\ci z\,|\,\emptyset\,,~  x\,\ci u\,|\,\emptyset\,,~  y\,\ci z\,|\,\emptyset\,,~  y\,\ci u\,|\,\emptyset\,,~  z\,\ci u\,|\,\emptyset\,,  & \\
& x\,\ci z\,|\,y\,,~  x\,\ci z\,|\,u\,,\quad   x\,\ci u\,|\,y\,,~  x\,\ci u\,|\,z\,, & \\
& y\,\ci z\,|\,x\,,~ y\,\ci z\,|\,u\,,\quad y\,\ci u\,|\,x\,,~  y\,\ci u\,|\,z\,,\quad   z\,\ci u\,|\,x\,,~  z\,\ci u\,|\,y\,, & \\
& x\,\ci z\,|\,yu\,,~  x\,\ci u\,|\,yz\,,~  y\,\ci z\,|\,xu\,,~  y\,\ci u\,|\,xz\,,~  z\,\ci u\,|\,xy\,.  &
\end{eqnarray*}
It has 6 permutably equivalent versions.

\paragraph{\fbox{II.}}
Put $N=\{x,y,z,u\}$ and define the density $p$ of a binary random vector:
\begin{center}
\begin{tabular}{lclc}
& $x~y~z~u$ &&\\
$p:$ & $(\0 ,\0 ,\0 ,\0 )$ &$\mapsto$ & $1/2$ \\
& $(\1 ,\1 ,\1 ,\0 )$ &$\mapsto$ & $1/2$
\end{tabular}
\quad with zero values for other configurations.
\end{center}
It provides (strong) probabilistic representation of a polymatroid over $N$ with
the following integer-valued rank function:
$$
h(S) ~:=~ \min\,\{\,|S\cap\{x,y,z\}|\,,\,1\,\}
\quad \mbox{for any $S\subseteq N$.}
$$
Note it corresponds to the uniform matroid of rank 1 over $\{x,y,z\}$.
The induced CI structure involves 18 (pairs of) elementary CI statements:
\begin{eqnarray*}
& x\,\ci u\,|\,\emptyset\,,~  y\,\ci u\,|\,\emptyset\,,~  z\,\ci u\,|\,\emptyset\,,  & \\
& x\,\ci y\,|\,z\,,\quad  x\,\ci z\,|\,y\,,\quad x\,\ci u\,|\,y\,,~ x\,\ci u\,|\,z\,,  & \\
& y\,\ci z\,|\,x\,,\quad  y\,\ci u\,|\,x\,,~ y\,\ci u\,|\,z\,,\quad  z\,\ci u\,|\,x\,,~  z\,\ci u\,|\,y\,, & \\
& x\,\ci y\,|\,zu\,,~ x\,\ci z\,|\,yu\,,~  x\,\ci u\,|\,yz\,,~  y\,\ci z\,|\,xu\,,~  y\,\ci u\,|\,xz\,,~  z\,\ci u\,|\,xy\,. &
\end{eqnarray*}
It has 4 permutably equivalent versions.

\paragraph{\fbox{III.}}
Put $N=\{x,y,z,u\}$ and define the density $p$ of a binary random vector:
\begin{center}
\begin{tabular}{lclc}
& $x~y~z~u$ &&\\
$p:$ & $(\0 ,\0 ,\0 ,\0 )$ &$\mapsto$ & $1/2$ \\
& $(\1 ,\1 ,\1 ,\1 )$ &$\mapsto$ & $1/2$
\end{tabular}
\quad with zero values for other configurations.
\end{center}
It provides (strong) probabilistic representation of a polymatroid over $N$ with
the following integer-valued rank function:
$$
h(S) ~:=~  \min\,\{\,|S|\,,\,1\,\}
\quad \mbox{for any $S\subseteq N$.}
$$
Note it corresponds to the uniform matroid of rank 1 over $N$.
The induced CI structure involves 18 (pairs of) elementary CI statements:
\begin{eqnarray*}
& x\,\ci y\,|\,z\,,~ x\,\ci y\,|\,u\,,\quad  x\,\ci z\,|\,y\,,~  x\,\ci z\,|\,u\,,\quad  x\,\ci u\,|\,y\,,~ x\,\ci u\,|\,z\,, & \\
& y\,\ci z\,|\,x\,,~ y\,\ci z\,|\,u\,,\quad  y\,\ci u\,|\,x\,,~ y\,\ci u\,|\,z\,,\quad  z\,\ci u\,|\,x\,,~  z\,\ci u\,|\,y\,, & \\
& x\,\ci y\,|\,zu\,,~ x\,\ci z\,|\,yu\,,~  x\,\ci u\,|\,yz\,,~  y\,\ci z\,|\,xu\,,~  y\,\ci u\,|\,xz\,,~  z\,\ci u\,|\,xy\,. &
\end{eqnarray*}
It has only 1 permutably equivalent version.

\paragraph{\fbox{IV.}} 
Put $N=\{x,y,z,u\}$ and define the density $p$ of a binary random vector:
\begin{center}
\begin{tabular}{lclc}
& $x~y~z~u$ &&\\
$p:$ & $(\0 ,\0 ,\0 ,\0 )$ &$\mapsto$ & $1/4$ \\
& $(\0 ,\1 ,\1 ,\0 )$ &$\mapsto$ & $1/4$ \\
& $(\1 ,\0 ,\1 ,\0 )$ &$\mapsto$ & $1/4$ \\
& $(\1 ,\1 ,\0 ,\0 )$ &$\mapsto$ & $1/4$
\end{tabular}
\quad with zero values for other configurations.
\end{center}
It provides (strong) probabilistic representation of a polymatroid over $N$ with
the following integer-valued rank function:
$$
h(S) ~:=~ \min\,\{\,|S\cap \{x,y,z\}|\,,\, 2\,\}
\quad \mbox{for any $S\subseteq N$.}
$$
Note it corresponds to the uniform matroid of rank 2 over $\{x,y,z\}$.
The induced CI structure involves 18 (pairs of) elementary CI statements:
\begin{eqnarray*}
& x\,\ci y\,|\,\emptyset\,,~  x\,\ci z\,|\,\emptyset\,,~ x\,\ci u\,|\,\emptyset\,,~  y\,\ci z\,|\,\emptyset\,,~  y\,\ci u\,|\,\emptyset\,,~  z\,\ci u\,|\,\emptyset\,,  & \\
& x\,\ci y\,|\,u\,,\quad  x\,\ci z\,|\,u\,,\quad  x\,\ci u\,|\,y\,,~ x\,\ci u\,|\,z\,, & \\
& y\,\ci z\,|\,u\,,\quad  y\,\ci u\,|\,x\,,~ y\,\ci u\,|\,z\,,\quad  z\,\ci u\,|\,x\,,~  z\,\ci u\,|\,y\,, & \\
& x\,\ci u\,|\,yz\,,\quad  y\,\ci u\,|\,xz\,,\quad  z\,\ci u\,|\,xy\,. &
\end{eqnarray*}
It has 4 permutably equivalent versions.

\paragraph{\fbox{V.}} 
Put $N=\{x,y,z,u\}$ and define the density $p$ of a binary random vector:
\begin{center}
\begin{tabular}{lclc}
& $x~y~z~u$ &&\\
$p:$ & $(\0 ,\0 ,\0 ,\0 )$ &$\mapsto$ & $1/8$ \\
& $(\0 ,\0 ,\1 ,\1 )$ &$\mapsto$ & $1/8$ \\
& $(\0 ,\1 ,\0 ,\1 )$ &$\mapsto$ & $1/8$ \\
& $(\0 ,\1 ,\1 ,\0 )$ &$\mapsto$ & $1/8$ \\
& $(\1 ,\0 ,\0 ,\1 )$ &$\mapsto$ & $1/8$ \\
& $(\1 ,\0 ,\1 ,\0 )$ &$\mapsto$ & $1/8$ \\
& $(\1 ,\1 ,\0 ,\0 )$ &$\mapsto$ & $1/8$ \\
& $(\1 ,\1 ,\1 ,\1 )$ &$\mapsto$ & $1/8$
\end{tabular}
\quad with zero values for other configurations.
\end{center}
It provides (strong) probabilistic representation of a polymatroid over $N$ with
the following integer-valued rank function:
$$
h(S) ~:=~ \min\,\{\,|S|\,,\,3\,\}
\quad \mbox{for any $S\subseteq N$.}
$$
Note it corresponds to the uniform matroid of rank 3 over $N$.
The induced CI structure involves 18 (pairs of) elementary CI statements:
\begin{eqnarray*}
& x\,\ci y\,|\,\emptyset\,,~  x\,\ci z\,|\,\emptyset\,,~ x\,\ci u\,|\,\emptyset\,,~  y\,\ci z\,|\,\emptyset\,,~  y\,\ci u\,|\,\emptyset\,,~  z\,\ci u\,|\,\emptyset\,,  & \\
& x\,\ci y\,|\,z\,,~ x\,\ci y\,|\,u\,,\quad  x\,\ci z\,|\,y\,,~  x\,\ci z\,|\,u\,,\quad  x\,\ci u\,|\,y\,,~ x\,\ci u\,|\,z\,, & \\
& y\,\ci z\,|\,x\,,~ y\,\ci z\,|\,u\,,\quad  y\,\ci u\,|\,x\,,~ y\,\ci u\,|\,z\,,\quad  z\,\ci u\,|\,x\,,~  z\,\ci u\,|\,y\,. &
\end{eqnarray*}
It has only 1 permutably equivalent version.

\paragraph{\fbox{VI.}}
Put $N=\{x,y,z,u\}$ and define the density $p$ of a binary random vector:
\begin{center}
\begin{tabular}{lclc}
& $x~y~z~u$ &&\\
$p:$ & $(\0 ,\0 ,\0 ,\0 )$ &$\mapsto$ & $1/4$ \\
& $(\0 ,\0 ,\1 ,\1 )$ &$\mapsto$ & $1/4$ \\
& $(\1 ,\1 ,\0 ,\1 )$ &$\mapsto$ & $1/4$ \\
& $(\1 ,\1 ,\1 ,\0 )$ &$\mapsto$ & $1/4$
\end{tabular}
\quad with zero values for other configurations.
\end{center}
It provides (strong) probabilistic representation of a polymatroid over $N$ with
the following integer-valued rank function:
$$
h(S) ~:=~
\left\{
\begin{array}{cl}
1 & ~~\mbox{if $S=\{x,y\}$,}\\
\min\,\{\,|S|\,,\,2\,\} & ~~\mbox{otherwise,}
\end{array}
\right.
\quad \mbox{for any $S\subseteq N$.}
$$
Note that it corresponds to a (non-uniform) matroid over $N$.
The induced CI structure involves 14 (pairs of) elementary CI statements:
\begin{eqnarray*}
& x\,\ci z\,|\,\emptyset\,,~ x\,\ci u\,|\,\emptyset\,,~  y\,\ci z\,|\,\emptyset\,,~  y\,\ci u\,|\,\emptyset\,,~  z\,\ci u\,|\,\emptyset\,,  & \\
& x\,\ci z\,|\,y\,,\quad  x\,\ci u\,|\,y\,,\quad  y\,\ci z\,|\,x\,,~ y\,\ci u\,|\,x\,, & \\
& x\,\ci y\,|\,zu\,,~ x\,\ci z\,|\,yu\,,~  x\,\ci u\,|\,yz\,,~  y\,\ci z\,|\,xu\,,~  y\,\ci u\,|\,xz\,. &
\end{eqnarray*}
It has 6 permutably equivalent versions.

\paragraph{\fbox{VII.}}
Put $N=\{x,y,z,u\}$ and define the density of a ternary random vector:
\begin{center}
\begin{tabular}{lclc}
& $x~y~z~u$ &&\\
$p:$ & $(\0 ,\0 ,\0 ,\0 )$ &$\mapsto$ & $1/9$ \\
& $(\0 ,\1 ,\1 ,\1 )$ &$\mapsto$ & $1/9$ \\
& $(\0 ,\2 ,\2 ,\2 )$ &$\mapsto$ & $1/9$ \\
& $(\1 ,\0 ,\1 ,\2 )$ &$\mapsto$ & $1/9$ \\
& $(\1 ,\1 ,\2 ,\0 )$ &$\mapsto$ & $1/9$ \\
& $(\1 ,\2 ,\0 ,\1 )$ &$\mapsto$ & $1/9$ \\
& $(\2 ,\0 ,\2 ,\1 )$ &$\mapsto$ & $1/9$ \\
& $(\2 ,\1 ,\0 ,\2 )$ &$\mapsto$ & $1/9$ \\
& $(\2 ,\2 ,\1 ,\0 )$ &$\mapsto$ & $1/9$
\end{tabular}
\quad with zero values for other configurations.
\end{center}
It provides (strong) probabilistic representation of a polymatroid over $N$ with
the following integer-valued rank function:
$$
h(S) ~:=~ \min\,\{\,|S|\,,\,2\,\}
\quad \mbox{for any $S\subseteq N$.}
$$
Note it corresponds to the uniform matroid of rank 2 over $N$.
The induced CI structure involves 12 (pairs of) elementary CI statements:
\begin{eqnarray*}
& x\,\ci y\,|\,\emptyset\,,~  x\,\ci z\,|\,\emptyset\,,~ x\,\ci u\,|\,\emptyset\,,~  y\,\ci z\,|\,\emptyset\,,~
y\,\ci u\,|\,\emptyset\,,~  z\,\ci u\,|\,\emptyset\,,  & \\
& x\,\ci y\,|\,zu\,,~ x\,\ci z\,|\,yu\,,~  x\,\ci u\,|\,yz\,,~  y\,\ci z\,|\,xu\,,~  y\,\ci u\,|\,xz\,,~  z\,\ci u\,|\,xy\,. &
\end{eqnarray*}
It has only 1 permutably equivalent version.

\paragraph{\fbox{VIII.}} 
Put $N=\{x,y,z,u\}$ and define the density $p$ of a random vector, whose
components are binary except for $\xi_{x}$ which has 4 allowed values:
\begin{center}
\begin{tabular}{lclc}
& $x~y~z~u$ &&\\
$p:$ & $(\0 ,\0 ,\0 ,\0 )$ &$\mapsto$ & $1/4$ \\
& $(\1 ,\0 ,\1 ,\1 )$ &$\mapsto$ & $1/4$ \\
& $(\2 ,\1 ,\0 ,\1 )$ &$\mapsto$ & $1/4$ \\
& $(\3 ,\1 ,\1 ,\0 )$ &$\mapsto$ & $1/4$
\end{tabular}
\quad with zero values for other configurations.
\end{center}
It provides (strong) probabilistic representation of a polymatroid over $N$ with
the following integer-valued rank function:
$$
h(S) ~:=~
\left\{
\begin{array}{cl}
2 & ~~\mbox{for $S=\{x\}$,}\\
\min\,\{\,|S|\,,\,2\,\} & ~~\mbox{if\, $S\neq \{x\}$,}
\end{array}
\right.
\quad \mbox{for $S\subseteq N$.}
$$
Note that it does not correspond to a matroid.
The induced CI structure involves 12 (pairs of) elementary CI statements:
\begin{eqnarray*}
& y\,\ci z\,|\,\emptyset\,,\quad y\,\ci u\,|\,\emptyset\,,\quad  z\,\ci u\,|\,\emptyset\,,  & \\
& y\,\ci z\,|\,x\,,\quad  y\,\ci u\,|\,x\,,\quad  z\,\ci u\,|\,x\,, & \\
& x\,\ci y\,|\,zu\,,~ x\,\ci z\,|\,yu\,,~  x\,\ci u\,|\,yz\,,~  y\,\ci z\,|\,xu\,,~  y\,\ci u\,|\,xz\,,~  z\,\ci u\,|\,xy\,. &
\end{eqnarray*}
It has 4 permutably equivalent versions.

\paragraph{\fbox{IX.}} 
Put $N=\{x,y,z,u\}$ and define the density $p$ of a random vector, whose
components are binary except for $\xi_{u}$ which has 4 allowed values:
\begin{center}
\begin{tabular}{lclc}
& $x~y~z~u$ &&\\
$p:$ & $(\0 ,\0 ,\0 ,\0 )$ &$\mapsto$ & $1/8$ \\
& $(\0 ,\0 ,\1 ,\1 )$ &$\mapsto$ & $1/8$ \\
& $(\0 ,\1 ,\0 ,\2 )$ &$\mapsto$ & $1/8$ \\
& $(\0 ,\1 ,\1 ,\3 )$ &$\mapsto$ & $1/8$ \\
& $(\1 ,\0 ,\0 ,\3 )$ &$\mapsto$ & $1/8$ \\
& $(\1 ,\0 ,\1 ,\2 )$ &$\mapsto$ & $1/8$ \\
& $(\1 ,\1 ,\0 ,\1 )$ &$\mapsto$ & $1/8$ \\
& $(\1 ,\1 ,\1 ,\0 )$ &$\mapsto$ & $1/8$
\end{tabular}
\quad with zero values for other configurations.
\end{center}
It provides (strong) probabilistic representation of a polymatroid over $N$ with
the following integer-valued rank function:
$$
h(S) ~:=~
\left\{
\begin{array}{cl}
   |S| & ~~\mbox{if $u\not\in S$,}\\
\min\,\{\,|S|+1\,,\,3\,\} & ~~\mbox{if $u\in S$,}
\end{array}
\right.
\quad \mbox{for $S\subseteq N$.}
$$
Note that it does not correspond to a matroid.
The induced CI structure involves 12 (pairs of) elementary CI statements:
\begin{eqnarray*}
& x\,\ci y\,|\,\emptyset\,,~  x\,\ci z\,|\,\emptyset\,,~ x\,\ci u\,|\,\emptyset\,,~  y\,\ci z\,|\,\emptyset\,,~
y\,\ci u\,|\,\emptyset\,,~  z\,\ci u\,|\,\emptyset\,,  & \\
& x\,\ci y\,|\,z\,,\quad  x\,\ci z\,|\,y\,,\quad  y\,\ci z\,|\,x\,, & \\
& x\,\ci y\,|\,zu\,,\quad x\,\ci z\,|\,yu\,,\quad  y\,\ci z\,|\,xu\,. &
\end{eqnarray*}
It has 4 permutably equivalent versions.

\section{Appendix: former versions of CI properties}\label{sec.former-CI}
This is to commemorate former occurrence of the properties from Corollary~\ref{cor.CI-implic}
in the literature. One of them was proved in \cite{Spo94} and two of them formulated
without proofs in \cite{Stu94IJGS}. The (original) proofs of all these properties
were then published in \cite{Mat95CPC,Mat99CPC}; nevertheless, the CI implications were not
formulated there as explicitly as in this paper and their proofs were encrypted as parts of
composite proofs for compound propositions. On the other hand, all the properties were
formulated explicitly in a conference contribution \cite{StuBoc94} (without proofs), two of them
in the form of open questions (= conjectures). Here we recall the (main) ideas of the original proofs
from \cite{Mat95CPC,Mat99CPC}.

\begin{itemize}
\item The implication (I:1) was proved in \cite[Proposition\,3.1,\,1st\,part]{Mat95CPC}
with $[X,Y,Z,U]=[1,2,3,4]$. The argument there was similar to the proof of the first conditional
information inequality [\,1cI\,]: the CI assumptions allow one to construct
a certain probability distribution on $ZU$ which dominates the marginal
distribution on $ZU$ and the non-negativity of the respective Kullback-Leibler divergence implies
its vanishing, meaning $Z\ci U$. The implication also appeared as (D.2) in \cite{StuBoc94}:
$[X,Y,Z,U]=[C,D,A,B]$.
\item The implication (I:2) first appeared as (B.1) in \cite[\S\,5]{Stu94IJGS}; it was also reported
in \cite{StuBoc94} as (B.1) with substitution $[X,Y,Z,U]=[C,D,B,A]$. The claim was then proved in
\cite[Proposition\,2.1,(1)]{Mat95CPC} with substitution $[X,Y,Z,U]=[1,2,3,4]$. The argument there
was the transformation to a wider framework of $\sigma$-algebras and applying a special CI property
within this framework (see Section~\ref{ssec.app.3rd}).
\item The implication (I:3) was proved in \cite[Proposition\,3.1,\,2nd\,part]{Mat95CPC}
with $[X,Y,Z,U]=[2,1,4,3]$. The argument there was some calculation with heedful
cancelation terms (allowed by CI assumptions) and using some alternative definitions of CI.
The property (I:3) also appeared as (D.3) in \cite{StuBoc94} with substitution $[X,Y,Z,U]=[B,D,A,C]$.
\item The implication (I:4) first appeared as (B.2) in \cite[\S\,5]{Stu94IJGS} and was also reported
in \cite{StuBoc94} as (B.2) with substitution $[X,Y,Z,U]=[B,D,A,C]$. The claim was proved in
\cite[Proposition\,2.1,(2)]{Mat95CPC} with $[X,Y,Z,U]=[2,1,4,3]$. The argument there
was the same as in case of (I:2).
\item The implication (I:5) was proved in \cite[Proposition\,2.1,(3)]{Mat95CPC} using
the same arguments as in case of (I:2); take $[X,Y,Z,U]=[2,1,4,3]$ there. The property
also appeared as (B.3) in \cite{StuBoc94} with $[X,Y,Z,U]=[B,D,A,C]$.
\item The implication (I:6) was proved in \cite[Proposition\,2.1,(4)]{Mat95CPC} using
the same arguments as in case of (I:2); take $[X,Y,Z,U]=[2,1,4,3]$ there.
The property also appeared as (B.4) in \cite{StuBoc94} with $[X,Y,Z,U]=[B,D,A,C]$.
\item The implication (I:7) was proved in \cite[Proposition\,3.1,\,3rd\,part]{Mat95CPC}
with $[X,Y,Z,U]=[2,1,4,3]$. The argument there was analogous as in case of (I:3).
The property appeared as (D.4) in \cite{StuBoc94}: $[X,Y,Z,U]=[B,D,A,C]$.
\item The implication (I:8) was proved in \cite[Proposition\,2.1,(8)]{Mat95CPC} using
the same arguments as in case of (I:2); take $[X,Y,Z,U]=[1,2,3,4]$.
The property also appeared as (B.8) in \cite{StuBoc94} with $[X,Y,Z,U]=[A,C,B,D]$.
\item The implication (I:9) was proved in \cite[Proposition\,3.1,\,4th\,part]{Mat95CPC}
with $[X,Y,Z,U]=[1,2,3,4]$. The argument there was similar to the one in case of (I:1):
the CI assumptions allow one to construct a probability distribution on $XYZ$ dominated by the marginal
distribution on $XYZ$ and the non-negativity of the respective Kullback-Leibler divergence forces that it
vanishes, which gives $X\ci Y\,|\,Z$. The property also appeared as (C.3) in \cite{StuBoc94}
with $[X,Y,Z,U]=[A,C,B,D]$.
\item The implication (I:10) was proved in \cite[Proposition\,2.1,(5)]{Mat95CPC} using
the same arguments as in case of (I:2); take $[X,Y,Z,U]=[1,2,3,4]$. The property
also appeared as (B.5) in \cite{StuBoc94} with $[X,Y,Z,U]=[A,C,B,D]$.
\item The implication (I:11) was proved in \cite[Proposition\,2.1,(6)]{Mat95CPC} using
the same arguments as in case of (I:2); take $[X,Y,Z,U]=[1,2,3,4]$.
The property also appeared as (B.6) in \cite{StuBoc94} with $[X,Y,Z,U]=[A,C,B,D]$.
\item The implication (I:12) was proved in \cite[Proposition\,2.1,(7)]{Mat95CPC} using
the same arguments as in case of (I:2); take $[X,Y,Z,U]=[1,2,3,4]$.
The property also appeared as (B.7) in \cite{StuBoc94} with $[X,Y,Z,U]=[A,C,B,D]$.
\item The implication (I:13) first appeared in \cite{Spo94} and its elegant proof there was
based on an equivalent definition of CI in terms of factorization of the density. The proof from
\cite[Proposition\,4.1,\,1st\,part]{Mat95CPC} was based on the same argument with $[X,Y,Z,U]=[1,2,3,4]$.
The implication (I:13) also appeared as (D.1) in \cite{StuBoc94} where $[X,Y,Z,U]=[A,B,C,D]$.
\item The implication (I:14) was proved in \cite[Proposition\,4.1,\,2nd\,part]{Mat95CPC}
with $[X,Y,Z,U]=[1,2,3,4]$; the argument there was the uniqueness principle for factorizable distributions
with coinciding marginals. The implication also appeared as (C.1) in \cite{StuBoc94} with $[X,Y,Z,U]=[A,B,C,D]$.
\item The implication (I:15) was derived in \cite[\S\,2]{Mat99CPC} as a consequence of
the inequality [\,1cI\,] from \cite[Proposition~2.1]{Mat99CPC} in the text
below that claim; take $[X,Y,Z,U]=[1,2,3,4]$. The property also appeared as an open
question a) in \cite{StuBoc94} with $[X,Y,Z,U]=[A,D,B,C]$.
\item The implication (I:16) was derived in \cite[\S\,2]{Mat99CPC} as a consequence of
the inequality [\,2cI\,] from \cite[Proposition~2.1]{Mat99CPC} in the text
below that claim; take $[X,Y,Z,U]=[1,2,3,4]$. The property
also appeared as an open question b) in \cite{StuBoc94} with $[X,Y,Z,U]=[A,D,B,C]$.
\item The implication (I:17) was proved in \cite[Proposition\,2.1,(9)]{Mat95CPC} using
the same arguments as in case of (I:2); take $[X,Y,Z,U]=[1,2,3,4]$.
The property also appeared as (B.9) in \cite{StuBoc94} with $[X,Y,Z,U]=[A,C,B,D]$.
\item The implication (I:18) was proved in \cite[Proposition\,2.1,(10)]{Mat95CPC} using
the same arguments as in case of (I:2); take $[X,Y,Z,U]=[1,2,3,4]$.
The property also appeared as (B.10) in \cite{StuBoc94} with $[X,Y,Z,U]=[A,C,B,D]$.
\item The implication (I:19) was proved in \cite[Proposition\,4.1,\,3rd\,part]{Mat95CPC}
with $[X,Y,Z,U]=[1,2,3,4]$; the argument there was the uniqueness principle for factorizable distributions
with coinciding supports. The implication also appeared as (C.2) in \cite{StuBoc94} with $[X,Y,Z,U]=[B,C,A,D]$.
\end{itemize}



\begin{thebibliography}{99}

\bibitem{AKNS20IEEE}
B.~Abo~Khamis, P.G.~~Kolaitis, H.Q.~Ngo, D.~Suciu:
\newblock{Decision problems in information theory.}
\newblock{In: Proceedings of 47th International Colloquium on Automata, Languages and Programming (ICALP 2020), pp. 106:1--106:20.} 

\bibitem{Bir95}
G.~Birkhoff:
\newblock{Lattice Theory.}
\newblock{Third edition, American Mathematical Society, Colloquium Publications 25, Providence 1995.}

\bibitem{BHLS10}
R.~Bouckaert, R.~Hemmecke, S.~Lindner, M.~Studen\'{y}:
\newblock{Efficient algorithms for conditional independence inference.}
\newblock{Journal of Machine Learning Research {\em 11} (2010), 3453--3479.}

\bibitem{CY02IEEE}
T.H.~Chen, R.W.~Yeung:
\newblock{On a relation between information inequalities and group theory.}
\newblock{IEEE Transactions on Information Theory {\em 48} (2002), 1992--1995.}

\bibitem{CK86}
I.~Csisz\'{a}r, J.~K\"{o}rner:
\newblock{Information Theory: Coding Theorems for Discrete Memoryless Systems.}
\newblock{Second impression, Acad\'{e}miai Kiad\'{o}, Budapest 1986.}

\bibitem{Daw79JRSS}
A.P.~Dawid:
\newblock{Conditional independence in statistical theory.}
\newblock{Journal of the Royal Statistical Society B {\em 41}(1) (1979), 1--31.}

\bibitem{Edm70}
J.~Edmonds:
\newblock{Submodular functions, matroids, and certain polyhedra.}
\newblock{In: Combinatorial Structures and their Applications, Gordon and Breach, 1970, pp.\  69--87.}

\bibitem{Fuj91}
S.~Fujishige:
\newblock{Submodular Functions and Optimization.}
\newblock{Second edition, Elsevier, Amsterdam 2005.}

\bibitem{Ing71conf}
A.W.~Ingleton:
\newblock{Conditions for representability and transversality of matroids.}
\newblock{In: Lecture Notes in Computer Science 211, Springer, 1971, pp. 62--67.}

\bibitem{KacRom13IEEE}
T.~Kaced, A.~Romashchenko:
\newblock{Conditional information inequalities for entropic and almost entropic points.}
\newblock{IEEE Transactions on Information Theory {\em 59}(11) (2013), 7149--7167.}

\bibitem{KRV18IEEE}
T.~Kaced, A.~Romashchenko, N.~Vereshchagin:
\newblock{A conditional information inequality and its combinatorial applications.}
\newblock{IEEE Transactions on Information Theory {\em 64}(5) (2018), 3610--3615.}

\bibitem{Lau96}
S.L.~Lauritzen:
\newblock{Graphical Models.}
\newblock{Clarendon Press, Oxford 1996.}

\bibitem{Loe55}
M.~Lo\'{e}ve:
\newblock{Probability Theory, Foundations, Random Sequences.}
\newblock{Van Nostrand, Toronto 1955.}

\bibitem{MMRV02}
K.~Makarychev, Y.~Makarychev, A.~Romashchenko, N.~Vereshchagin:
\newblock{A new class of non-Shannon-type inequalities for entropies.}
\newblock{Communications in Information and Systems {\em 2}(2) (2002), 147--166.}

\bibitem{Mat92conf}
F.~Mat\'{u}\v{s}:
\newblock{Ascending and descending conditional independence relations.}
\newblock{In: Transactions of the 11th Prague Conference on Information Theory, Statistical Decision
Functions and Random Processes, volume B, Academia, Prague 1992, pp. 189--200.}

\bibitem{Mat94TCS}
F.~Mat\'{u}\v{s}:
\newblock{Stochastic independence, algebraic independence and abstract connectedness.}
\newblock{Theoretical Computer Science {\em 134} (1994), 455--471.}

\bibitem{Mat94IJGS}
F.~Mat\'{u}\v{s}:
\newblock{Probabilistic conditional independence structures and matroid theory: background.}
\newblock{International Journal of General Systems {\em 22} (1994), 185--196.}

\bibitem{Mat94DM}
F.~Mat\'{u}\v{s}:
\newblock{Extreme convex set functions with many non-negative differences.}
\newblock{Discrete Mathematics {\em 135} (1994), 177--191.}

\bibitem{MatStu95CPC}
F.~Mat\'{u}\v{s} and M.~Studen\'{y}:
\newblock{Conditional independences among four random variables I.}
\newblock{Combinatorics, Probability and Computing {\em 4} (1995), 269--278.}

\bibitem{Mat95CPC}
F.~Mat\'{u}\v{s}:
\newblock{Conditional independences among four random variables II.}
\newblock{Combinatorics, Probability and Computing {\em 4} (1995), 407--417.}

\bibitem{Mat99CPC}
F.~Mat\'{u}\v{s}:
\newblock{Conditional independences among four random variables III: final conclusion.}
\newblock{Combinatorics, Probability and Computing {\em 8} (1999), 269--276.}

\bibitem{Mat07IEEE}
F.~Mat\'{u}\v{s}:
\newblock{Two constructions on limits of entropy functions.}
\newblock{IEEE Transactions on Information Theory {\em 53} (2007) 320--330.}

\bibitem{Mat07ISIT}
F.~Mat\'{u}\v{s}:
\newblock{Infinitely many information inequalities.}
\newblock{In: Proceedigns of IEEE International Symposium on Information Theory, Nice 2007, pp.\ 41-44.}

\bibitem{MouRol84S}
M.~Mouchart, J.M.~Rolin:
\newblock{A note on conditional independence with statistical applications.}
\newblock{Statistica {\em 44} (1984), 557--584.}

\bibitem{Oxl11}
J.G.~Oxley:
\newblock{Matroid Theory.}
\newblock{Second edition, Oxford University Press, New York 2011.}

\bibitem{Pad12}
C.~Padro:
\newblock{Lecture Notes in Secrete Sharing.}
\newblock{Cryptology E-print Archive, Report 2012/674.}

\bibitem{Pea88}
J.~Pearl:
\newblock{Probabilistic Reasoning in Intelligent Systems: Networks of Plausible Inference.}
\newblock{Morgan Kaufmann, San Francisco 1988.}

\bibitem{Pip86}
N.~Pippenger:
\newblock{What are the laws of information theory.}
\newblock{In: 1986 Special Problems on Communication and Computation Conference, pp.\ 3--5.}

\bibitem{PRW08}
A.~Postnikov, V.~Reiner, L.~Williams:
\newblock{Faces of generalized permutohedra.}
\newblock{Documenta Mathematica {\em 13} (2008), 207--273.}

\bibitem{Sch03}
A.~Schrijver:
\newblock{Combinatorial Optimization: Polyhedra and Efficiency, volume B.}
\newblock{Springer, 2003.}

\bibitem{Sim07}
P.~\v{S}ime\v{c}ek:
\newblock{Independence models (in Czech).}
\newblock{PhD thesis, Charles University, Prague, Czech Republic, 2007.}

\bibitem{Spo80JPL}
W.~Spohn:
\newblock{Stochastic independence, causal independence and shieldability.}
\newblock{Journal of Philosophical Logic {\em 9}(1) (1980), 73--99.}

\bibitem{Spo94}
W.~Spohn:
\newblock{On the properties of conditional independence.}
\newblock{In: Patrick Suppes: Scientific Philosopher (P.\, Humphreys ed.), Kluwer, Dordrecht 1994, pp.\ 173--196.}

\bibitem{Stu89}
M.~Studen\'{y}:
\newblock{Multiinformation and the problem of characterization of conditional independence relations.}
\newblock{Problems of Control and Information Theory {\em 18}(1) (1989), 3--16.}

\bibitem{Stu92conf}
M.~Studen\'{y}:
\newblock{Conditional independence relations have no finite complete characterization.}
\newblock{In: Transactions of the 11th Prague Conference on Information Theory,
Statistical Decision Functions and Random Processes, volume B, Academia, Prague 1992, pp. 377--396.}

\bibitem{Stu94IJGS}
M.~Studen\'{y}:
\newblock{Structural semigraphoids.}
\newblock{International Journal of General Systems {\em 22} (1994), 207--217.}

\bibitem{StuBoc94}
M.~Studen\'{y}, P.~Bo\v{c}ek:
\newblock{CI-models arising among 4 random variables.}
\newblock{In: Proceedings of WUPES`94, September 11-15, 1994,  pp.\ 268-282.}

\bibitem{Stu05}
M.~Studen\'{y}:
\newblock{Probabilistic Conditional Independence Structures.}
\newblock{Springer, London 2005.}

\bibitem{SK16DAM}
M.~Studen\'{y}, T.~Kroupa:
\newblock{Core-based criterion for extreme supermodular functions.}
\newblock{Discrete Applied Mathematics {\em 206} (2016), 122--151.}

\bibitem{Whi35AJM}
H.~Whitney:
\newblock{On the abstract properties of linear dependence.}
\newblock{American Journal of Mathematics {\em 57}(3) (1935), 509--533.}

\bibitem{Yeu08}
R.W.~Yeung:
\newblock{Information Theory and Network Coding.}
\newblock{Springer, New York 2008.}

\bibitem{ZhaYeu97ITI}
Z.~Zhang, R.W.~Yeung:
\newblock{A non-Shannon-type conditional inequality of information quantities.}
\newblock{IEEE Transactions on Information Theory {\em 43} (1997), 1982--1986.}

\bibitem{ZhaYeu98ITI}
Z.~Zhang, R.W.~Yeung:
\newblock{On characterization of entropy function via information inequalities.}
\newblock{IEEE Transactions on Information Theory {\em 44} (1998), 1440--1450.}

\bibitem{Zie95}
G.M.~Ziegler:
\newblock{Lectures on Polytopes.}
\newblock{Springer, New York 1995.}

\end{thebibliography}
\end{document}